\documentclass[fleqn,usenatbib]{mnras}

\usepackage{newtxtext,newtxmath}
\usepackage[T1]{fontenc}

\DeclareRobustCommand{\VAN}[3]{#2}
\let\VANthebibliography\thebibliography
\def\thebibliography{\DeclareRobustCommand{\VAN}[3]{##3}\VANthebibliography}

\usepackage{graphicx}	
\usepackage{amsmath}	
\usepackage{subcaption}
\usepackage{booktabs}
\usepackage{tabularx,array} 
\newcolumntype{Z}{>{\centering\arraybackslash}p{1.2cm}}

\newcommand{\solar}{$M_\odot$}
\newcommand{\maxi}{\textit{MAXI}}
\newcommand{\swift}{\textit{Swift}}
\newcommand{\rxte}{\textit{RXTE}}

\newcommand{\pcm}{\,cm$^{-2}$}	

\title[The galactic BH-LMXB distribution]{A new independent look at the galactic black hole low-mass X-ray binary distribution}

\author[Abdulghani et al.]{
Y. Abdulghani,$^{1}$\thanks{E-mail: youssefabdulghani@montana.edu}
A. M. Lohfink,$^{1}$
J. Chauhan$^{2, 1}$
\\
$^{1}$Department of Physics, Montana State University, P.O. Box 173840, Bozeman, MT 59717-3840, USA\\
$^{2}$School of Physics and Astronomy, University of Leicester, University Rd, Leicester LE1 7RH, United Kingdom\\
}

\date{Accepted XXX. Received YYY; in original form ZZZ}

\pubyear{2015}

\begin{document}
\label{firstpage}
\pagerange{\pageref{firstpage}--\pageref{lastpage}}
\maketitle

\begin{abstract}
Investigations of the Galactic black hole low-mass X-ray binaries (BH-LMXBs) offer valuable insights into the elusive black hole population in the Milky Way. Motivated by recent tensions in the natal kick velocity distribution and BH mass distribution of BH-LMXBs, we revisit the spatial distribution of the Galactic BH-LMXBs using a new set of distance measurements obtained from an X-ray spectral modelling framework that we introduced in earlier work. We perform a multiparameter simulation study to mitigate part of the bias present in our prior estimates and gain insights into possible observational selection effects that affect the observed population. We derive a bias correction factor, described by a Pareto probability density function that decays slowly at low distances and then follows an inverse-square law dependence on distance in the high distances limit. We then construct a bias-corrected, literature-independent, Galactic spatial distribution that clearly traces spiral arm structures and shows a deficit of sources very close to the Galactic centre, which might be explained due to high extinction or a true paucity of these sources at that region. Further analysis of the simulation results provides hints for a hidden population of BH‐LMXBs at low Galactic heights. Lastly, we estimate the root-mean-squared Galactic height and find that it is most compatible with a hybrid scenario of BH formation, with some BHs receiving high natal kicks and thus propelled further from the thin disc plane while others receiving low natal kicks and remaining close to their birthplace.
\end{abstract}

\begin{keywords} X-rays: binaries – stars: black holes – stars: distances – accretion, accretion discs – methods: statistical - Galaxy: general
\end{keywords}



\section{Introduction}
Low-mass black hole X-ray binaries (BH-LMXBs) are systems in which a low-mass companion star ($\lesssim$ 1\solar) orbits a black hole (BH). The black hole gains mass from the companion star via Roche lobe overflow, and the accreted material forms an accretion disc \citep{shakura1973}. These systems are predominantly detected during transient outburst events when the X-ray flux increases significantly \cite[e.g.][]{Remillard2006,Tetarenko2016}. Recent increased interest and telescopes capability has contributed to the enhanced detection of these transient X-ray binaries within our Galaxy by providing greater sensitivity and expanding sky coverage \citep{Tetarenko2016}. Despite the improved observational power, constraining the nature of the BH-LMXB population continues to present a formidable challenge.

Currently, the number of black hole (BH) remnants in the Milky Way is estimated to be on the order of $\sim 10^8$–$10^9$ \citep[e.g.,][]{Brown1994,Timmes1996,Samland1998}. Some of these BHs reside in high-mass X-ray binaries (BH-HMXBs), which are characterized by massive stellar companions with masses $\gtrsim 10$ $M_\odot$ \citep{MacLeod2023}. However, BH-HMXBs are observed relatively infrequently, likely due to the short lifespans of massive stars and the complex binary evolution pathways they undergo, such as the stability of mass transfer and the system's ability to survive common-envelope phases \citep{MacLeod2023}. While BH-HMXBs are valuable for understanding specific evolutionary channels, their low-mass X-ray binaries counterparts (BH-LMXBs) offer a larger observable sample and fewer detection biases. Their evolutionary histories also pose important theoretical challenges. As a result, BH-LMXBs provide a more accessible window into the broader black hole population and formation mechanisms in the Galaxy. In this work, we focus on studying the BH-LMXB population to gain deeper insight into Galactic BH demographics.

The formation of BH-LMXBs is a complex issue involving stellar evolution, dynamical modeling, and observational data \citep[e.g.][]{Wang2016}. Several formation channels for BH X-ray binaries have been proposed \citep[see review in][]{MacLeod2023}, mostly based on well-understood binary systems. The standard model involves common envelope (CE) evolution in an isolated binary, where a massive star expands, enveloping both stars. The secondary can spiral inward, ejecting the common envelope and leading to a BH-LMXB, if the primary becomes a BH, certain dynamical conditions are met, and the secondary is roughly 1 \solar~\citep[e.g.][]{Ivanova2013}. However, this model has been challenged, as the secondary low mass makes binary survival after the CE unlikely \citep{Podsiadlowski2003,Naoz2016}.

Two alternative channels involving tidal capture are also considered \citep{MacLeod2023}. One occurs when a wide binary dynamically encounters the primary, with the BH already formed before the capture, avoiding post-CE issues. This requires a dense environment like globular clusters or the Galactic Center \citep{MacLeod2023}. The other involves a hierarchical triple system, where a tertiary's dynamical interactions bring the secondary closer to the primary (the BH), initiating mass transfer. Though the CE phase may still happen in this case, three-body dynamics improve the chances of survival, addressing the standard scenario's challenges \citep{Naoz2016,Shariat2024}. A supermassive black hole, like the one in the Galactic Center, could act as the tertiary \citep{Lu2019,Hoang2022}.

Regardless of the channels, the transformation of the primary from a giant to a compact star can also follow different paths; it can be accompanied by a symmetric supernova explosion, asymmetric supernova explosion or can happen with minimal mass loss, i.e an implosion \citep[e.g.][]{fryer2001,Heger2023,fryer2012}. These three possibilities have significant implications for the velocity that the system may gain during the birth of the compact primary. This velocity has been widely referred to as the ``natal velocity'' or the ``natal kick velocity''. Furthermore, the core of the primary can either collapse directly into a BH or first into a proto-neutron star and subsequently into a BH due to mass fallback \citep{fryer2001}.

Studies of the peculiar motion of a large number of neutron star LMXBs have led to strong evidence that neutron star LMXBs receive large natal kicks \citep[e.g.][]{Hobbs2005,Faucher2006}. Several attempts have been made over the past two decades to determine whether BH-LMXBs receive equally large natal kicks \citep{Jonker2004,Repetto2012,Mandel2016,Repetto2017,Atri2019}. At present, low natal kick and high natal kick distributions have almost the same degree of evidence, and tensions to reconcile both scenarios are high \citep{Mandel2016,MacLeod2023,Nagarajan2024}. Nevertheless, most recently, there has been an appeal to accept a bi-modality in the natal kick distribution for BH-LMXB \citep{Nagarajan2024}. Especially with the latest, very high-confidence discovery of a tertiary and low natal kick in the BH-LMXB system: V404 Cygni \citep{Burdge2024}.

The efficiency of a BH-LMXB formation channel is strongly influenced by the magnitude of the natal kick received by the system \citep[e.g.][]{Shariat2024}. High natal kick velocities generally act to reduce the overall efficiency of X-ray binary formation by increasing the likelihood of binary disruption or by altering orbits to be less favorable for subsequent mass transfer. For example, the hierarchical triple formation channel generally favors a low natal velocity, as higher velocities often lead to the secondary detaching from the tertiary, thereby preventing the subsequent inward spiral \citep{Naoz2016, Burdge2024, Shariat2024}. Likewise, high natal velocities can significantly hinder the efficiency of the wide binary dynamical interaction channel, as a high-velocity BH (or neutron star) would quickly pass by a slower potential companion, preventing the energy dissipation required for tidal capture \citep{MacLeod2023}.

On a related note, discrepancies between the distribution of BH masses deduced from electromagnetic observations and those inferred from gravitational wave observations \citep{Fishbach2022, Siegel2023} raise additional questions. One proposed solution to address this tension is to demonstrate that two distinct populations of BHs exist, both in turn influenced by a combination of observational and astrophysical selection effects \citep{Fishbach2022, Siegel2023}. \cite{Jonker2021} pointed out that observational selection effects might explain the bias against massive BHs in the observed BH-LMXBs mass distribution. Astrophysical selection effects were also suggested to be one of the reasons behind the observed lower BH mass gap within the 2-5 \solar~range in the BH-LMXB mass distributions \citep{Ozel2010,Farr2011,Kriedberg2012,Siegel2023}. These selection effects could be explained in the context of natal kicks. More specifically, since mass influences natal kick velocity due to the conservation of momentum, systems with high-mass BHs and low natal velocities will be more challenging to identify observationally. Conversely, large natal kicks for low-mass BHs can disrupt the system, reducing the likelihood of BH-LMXB formation. This could introduce a selection effect where BH-LMXBs predominantly form around sufficiently massive BHs. However, it remains unclear how this explanation aligns with neutron star LMXBs, which have been shown to receive high natal kicks \citep[e.g.][]{Hobbs2005}, despite the assumption that the formation processes of BH and neutron star LMXBs should not differ significantly. One possible reconciliation is that if the natal kick occurs on a very short timescale ($\sim$tens of milliseconds), it could expel mass from the system before the proto-neutron star has time to accrete enough material to collapse into a BH \citep[e.g.][]{Fryer2022}. However, if the natal kick occurs over a longer timescale ($\sim$hundreds of milliseconds), fallback accretion could still occur, leading to BH formation. This suggests that the timescale of the natal kick plays a crucial role in determining whether a compact object remains a neutron star or collapses into a BH, but further studies are needed to fully explain the observed differences \citep{Jonker2021}.

Therefore, studies of the Milky Way's BH population are critically important in light of the recent tensions in both the natal kick distribution and the BH mass distribution. Throughout the past two decades, there have been several investigations into the population properties of BH-LMXBs in the Milky Way \citep[e.g.][]{Pfahl2003,Jonker2004,Ozel2010,Vanhaaften2015,Repetto2017,Gandhi2020,Siegel2023,Shikauchi2023}. However, past studies were inherently hampered by very small sample sizes and poor distance estimations, which were usually obtained from the existing literature. 

Motivated by this lack of robust BH distribution estimates, we aim to provide a new independent examination of the Galactic spatial distribution of a relatively large sample of the observed BH-LMXBs population.
Our work builds on the findings from prior work presented in \citet[][herein A24]{Abdulghani2024}, where we developed a Bayesian framework to find dependable estimations of BH-LMXBs distance using the soft state and the soft-to-hard transition X-ray spectra as they are observed during an X-ray binary outburst (temporary enhanced accretion event).

The distance estimation method presented by A24 determines the distances to BH-LMXBs by analyzing their X-ray spectra during outbursts, using two distinct spectral phases. The first phase corresponds to a disc-dominated state, where the X-ray emission arises primarily from the accretion disc around the black hole. This emission is modeled using a multicolored disc blackbody \citep{Mitsuda1984, Makishima1986}, where the disc normalization depends on the distance to the source, the inner disc radius, the system's inclination, the black hole's spin, and its mass. The second phase occurs during the characteristic spectral transition, in which the system evolves from the soft, disc-dominated state to a harder state with more prominent high-energy X-ray emission. Empirical studies \citep[e.g.,][]{Maccarone2003, Vahdat2019} have shown that this transition typically occurs at a relatively well-constrained fraction of the Eddington luminosity, which is itself a function of black hole mass. Measuring the X-ray flux at this transition thus provides a complementary constraint on the distance, through the relation ${\rm F} = \frac{\rm L}{4\pi {\rm D}^2}$, where $\rm F$ is the flux, $\rm L$ the luminosity, and $\rm D$  the distance to the source. By combining constraints from both the disc-dominated phase and the transition point, the A24 framework enables a robust and independent estimation of the distance to BH-LMXBs.

Various observational and intrinsic properties can bias these distance estimates, which in turn could affect the spatial distribution of the observed BH-LMXB population. We thus first need to conduct extensive multi-parameter simulations that involve varying the properties and observational parameters of a synthetic BH-LMXB so that we can arrive at the characteristics of the true BH population. 

This article is structured as follows: section~\ref{sec:simulation-study} introduces the BH-LMXB simulation method and its results. Next, section~\ref{sec:detailedobservationalview} provides the corrected spatial distributions of the BH-LMXB in our Galaxy. We then discuss the implications of the simulations and the corrected observed distribution in section~\ref{sec:discussion}. Lastly, we summarise our findings and their implications in section~\ref{sec:conclusions}.

\section{Observational Bias Effects on A24's Soft-State Distances: Simulation Study}
\label{sec:simulation-study}

To investigate the observational effects on the distances, we simulate synthetic X-ray spectra based on different observational and intrinsic properties of a hypothetical BH-LMXB that we assume is in the soft state during an outburst. We then model the simulated spectrum with a model appropriate for the soft state and obtain a best-fit as well as statistical uncertainties on the best-fit model parameters. The obtained best-fit spectral parameters are then used to constrain the source's distance using the method developed in A24. Subsequently, by comparing the estimated distance to the distance assumed when simulating the spectrum, we can then assess the uncertainties and biases in the estimated distances arising from a particular combination of BH-LMXB properties. The resulting distance estimates will also be used to improve our understanding of the observational effects on the observed BH-LMXB population in A24.

\subsection{Simulation methodology}

The spectral simulations in this paper are based on the fact that the observed X-ray spectrum of a BH-LMXB can be adequately modelled to first order, with only two components that are modified by interstellar absorption; up-scattered hard X-ray emission from the hot corona believed to be present in the inner regions near the BH \citep{haardt1991, Haardt1993, Kara2016}, and thermal emission from accretion disc \citep{shakura1973}. While the soft state spectra of BH-LMXBs are dominated by the accretion disc component \citep{shakura1973}, there often is also a weak continuum component present, but its strength can vary \citep{haardt1991, Haardt1993}. In our simulations, we, therefore, vary the relative strength of the two components; this is captured in the disc-to-total ratio. In the remainder of this section, we describe the detailed method used to generate synthetic spectra of possible realisations of an observed X-ray spectrum of the soft state. 

\begin{table}
\centering
\caption{This table shows the parameter value set used in the spectral simulations.}
\begin{tabular}{cc}
\toprule
\textbf{Parameter (units)} & \textbf{Values} \\
\midrule
$N_{\text{H}}$ ($10^{22}$ cm$^{-2}$)       & 0.1, 0.5, 5, 10 \\
$D_{\text{in}}$ (kpc)                       & 1, 2, 3, 4, 5, 6, 8, 12, 18, 26 \\
$\Gamma$                                  & 1.7, 2, 3 \\
$T$ (keV)                                 & 0.5, 0.7, 1 \\
$a$                                       & 0.0, 0.998 \\
$M$ ($M_{\odot}$)                         & 6, 8, 10 \\
$i$ (\textdegree)                          & 0, 60, 80 \\
disc-to-total ratio                       & 0.2, 0.5, 0.9 \\
\maxi~Exposures (sec)                      & 400, 1500, 10000 \\
\swift~Exposures (sec)                     & 400, 1000, 5000 \\
\bottomrule
\end{tabular}
\label{tab:sim_param}
\end{table}

We simulated the BH-LMXB spectra in \texttt{XSPEC}~\citep[v12.14.1;][]{xspec_ref} using the model \texttt{TBabs(powerlaw+ezdiskbb)}. By considering different realistic combinations of the spectral model and observational parameters (Table~\ref{tab:sim_param}), we can investigate their effect on BH-LMXB distance estimates and detection. 

The disc component, modelled by \texttt{ezdiskbb} \citep{zimmerman2005}, is fully determined by the maximum temperature \(T\) of the accretion disc and its normalisation. The normalisation itself is established through the distance $D$, inclination $i$, and intrinsic system properties (BH mass and BH spin). More explicitly, the normalisation can be parametrised in the following way:
\begin{equation}
    N_{\texttt{ezdiskbb}} = \left( \frac{r_{\text{in}} \text{ [km]}}{f_{\text{col}}^2 \left[D/10 \text{ kpc}\right]} \right)^2 \cos i~\Upsilon(i)~g_{GR}(a,i)~g_{NT}(a)
    \label{eq:norm_ezdiscbb_corrections}
\end{equation}
where the $\Upsilon(i)$, $g_{GR}(a,i)$, and $g_{NT}(a)$ are correction factors based on BH spin and the inclination angle of the disc (see section 5.2 in A24 for more details). The color correction factor (or spectral hardening factor) $f_{\text{col}}$ we assume to be approximately 1.7. The disc inner radius $r_{\text{in}}$, we assume, is at the ISCO, so it is a function of $a$ and $M$ and can be written in this form:
\begin{equation}
r_{\text{in}} = \frac{\zeta(a)GM}{c^2}.
    \label{eq:r_in}
\end{equation}
where, for our purposes, $\zeta(a)$ is a multiplier that depends on the value of BH spin $a$ (e.g $\zeta(0) = 6$). 

A simple power law is used to model the up-scattered high-energy emission. This power law is characterized by two parameters: the photon index \(\Gamma\) and a normalisation factor that determines the flux. In our simulations, we assume that the power law flux represents the remainder of the total unabsorbed flux after setting a specific disc-to-total flux ratio. Additionally, to account for possible different interstellar absorption columns along the line of sight to our simulated black hole X-ray binary in outburst, we vary the column density, \(N_{\text{H}}\) parameter, of the \texttt{TBabs} model \citep{wilms2000}.

Consequently, our simulations involve selecting specific values for \(\Gamma\), the disc-to-total flux ratio, the maximum disc temperature \(T\), the inclination \(i\), the BH mass \(M\), and the BH spin \(a\). Spectra are then simulated for all possible combinations of these parameters. One of our main goals from the simulations is to correct the biases intrinsic to the instruments, which are present in the A24 distance estimations. So, the parameter values considered in the simulations were chosen primarily based on the observed ranges in the A24 spectral modelling results, and the priors on BH properties we used in A24. We selected three \(\Gamma\) values, 1.7, 2, and 3, since typical soft-state values lie between \(\Gamma = 2\) and \(\Gamma = 2.5\) \citep{Remillard2006}; the \(\Gamma = 1.7\) case accounts for a harder-than-usual spectrum, while \(\Gamma = 3\) represents a softer-than-usual scenario. Moreover, we chose three values for the maximum disc temperature, \(T = 0.5\), 0.7, and 1 keV \citep[e.g.][]{munoz-darias2013}, two values for the BH spin, no-spin \(a=0\), and maximal-spin \(a=0.998\) \citep{reynolds2013}, and three values for the BH mass, \(M = 6\), 8, and 10~\(\mathrm{M}_{\odot}\) \citep{Ozel2010}. We also considered three values for the disc-to-total flux ratio, 0.2, 0.5, and 0.9 \citep{Remillard2006} and selected three disc inclination values that span the minimum ($i=0$\textdegree), expectation value ($i=60$\textdegree) and near the possible maximum ($i=80$\textdegree). We adopted three values for each instrument for the exposure time, based on the values observed when analysing the A24 sample.

In addition to these parameters, we implemented non-uniform grids for the distance and \(N_{\text{H}}\) values. Specifically, we constructed a 4-by-10 \(D_{\text{in}}-N_{\text{H}}\) grid that covers distances likely to occur within the Galactic sample and spans the typical \(N_{\text{H}}\) values up to the “maximum” value \citep{HI4PI2016}. Table~\ref{tab:sim_param} presents the full set of simulation parameters for each instrument.

The spectra simulation were performed using the \textit{fakeit} command in \texttt{XSPEC}. Spectra are generated based on the predetermined set of model parameters. The simulation procedure was conducted using representative background and response files from both Neil Gehrels Swift Observatory/X-ray Telescope - Windowed Timing mode \citep[\swift/XRT-WT data;][]{xrt} and the Monitor of All-sky X-ray Image/Gas Slit Counter \citep[\maxi/GSC;][]{maxi}. We binned the simulated spectra to a signal-to-noise ratio (SNR) of 3, consistent with the data reduction methodology used in A24. We do not take into account in our simulations the \swift/XRT pile-up effects \citep{Mineo2006} that affect sources with high flux ($\gtrsim$100\,counts\,s$^{-1}$) \citep{Romano2006}, since we assume that the simulated spectrum would be the resulting spectrum after correcting for pile-up.

The generated spectra have random Poisson errors, to ensure that we get a representative sample of the simulated spectra, each simulation was repeated 300 times. However, to provide a concise synopsis of the modelling results of the 300 simulated spectra per parameter combination, we calculated the weighted median of the estimated distances after generating the 300 spectra. The weight is constructed from the mean fractional uncertainty by utilising lower and upper 90\% confidence bounds of the \texttt{ezdiskbb} normalisation parameter. The 90\% confidence bounds of the \texttt{ezdiskbb} normalisation for each simulated spectrum were calculated using \texttt{XSPEC}'s \textit{error} command. 

The mean fractional uncertainty is thus taken to be $f_{\delta} = (D_\text{est,upper}-D_\text{est,lower})/D_\text{est,best-fit}$ so that the weight is $w=1/f_{\delta}$. Moreover, the weighted median of the 300 simulations was taken to be the quoted estimated distance $D_{\text{est}}$ for that parameter combination.

We also evaluated whether a source was too faint at certain parameter combinations to obtain a distance estimate via this spectral method. Specifically, if the simulated spectrum is of such low quality that the number of data points in the binned (SNR~$\geq$~3) spectrum is less than the number of fit parameters, we consider the source ``undetectable" for that combination.

In summary, the simulation procedure was as follows:
\begin{enumerate}
    \item For each grid point in the $4\times10$ $D_{\text{in}}$--$N_{\text{H}}$ grid, simulate 300 spectra with random error
    \item Repeat step 1 for each of the 1,458 combinations of the other parameters listed in Table~\ref{tab:sim_param}.  This results in a total of $17,496,000$ simulated spectra per instrument for each input parameter combination
    \item Reduce these $17,496,000$ spectra per instrument to $58,320$ by computing the weighted median (as described above) of the 300 repeated simulations for each combination
    \item Use these $58,320$ combinations per instrument as the starting point for all subsequent results presented hereafter 
\end{enumerate}

\subsection{Simulation results}

A total of 34,992,000 simulations were generated. The computations were performed on the MSU-Tempest HPC research cluster\footnote{\url{https://www.montana.edu/uit/rci/tempest/}} and consumed around 17,000 CPU hours. Since the results of the simulations are extremely large, it is intractable to look at every possible simulation parameter combination. Therefore, we focus our investigation on three avenues. We begin by examining how the deviation from the input distance varies as a function of the input distance for each of the eight simulation parameters (Section \ref{sec:sim_ind}). We then probe important pairwise parameter interactions through a Bayesian Information Criterion (BIC) selection process (Section \ref{sec:pairwise}). Finally, we provide a holistic view of the estimated distance distributions, to utilise these distributions as bias correctors. The scripts used to generate the simulations and the distilled results are available in a public repository\footnote{https://github.com/ysabdulghani/xrb-population}.

\subsubsection{Effects of individual parameters}\label{sec:sim_ind}

\begin{figure*}
    \centering
    \begin{subfigure}[t]{0.49\textwidth}
        \includegraphics[width=\textwidth]{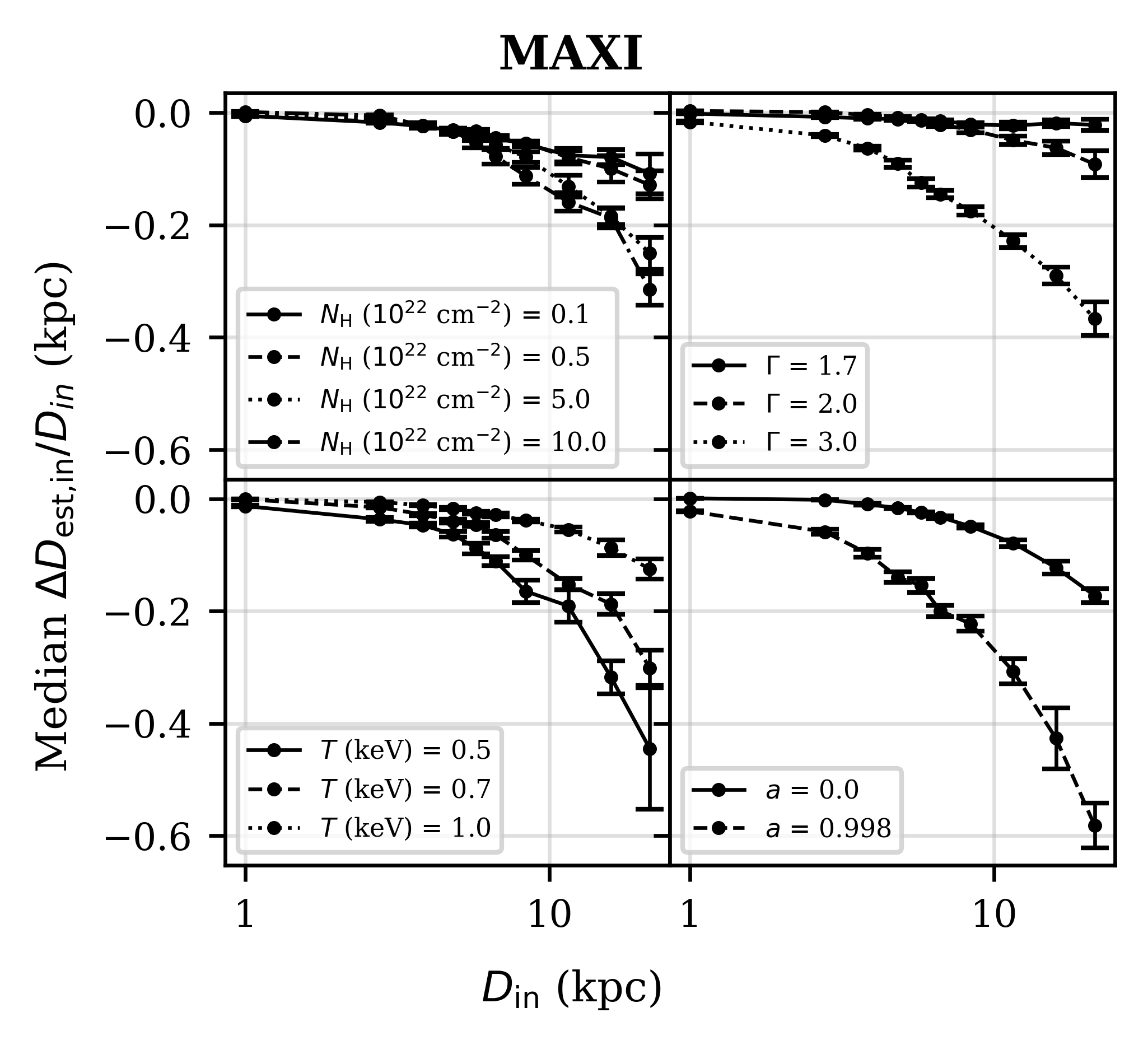}
        \label{fig:trendline1-maxi}
    \end{subfigure}
    \hfill
    \begin{subfigure}[t]{0.49\textwidth}
        \includegraphics[width=\textwidth]{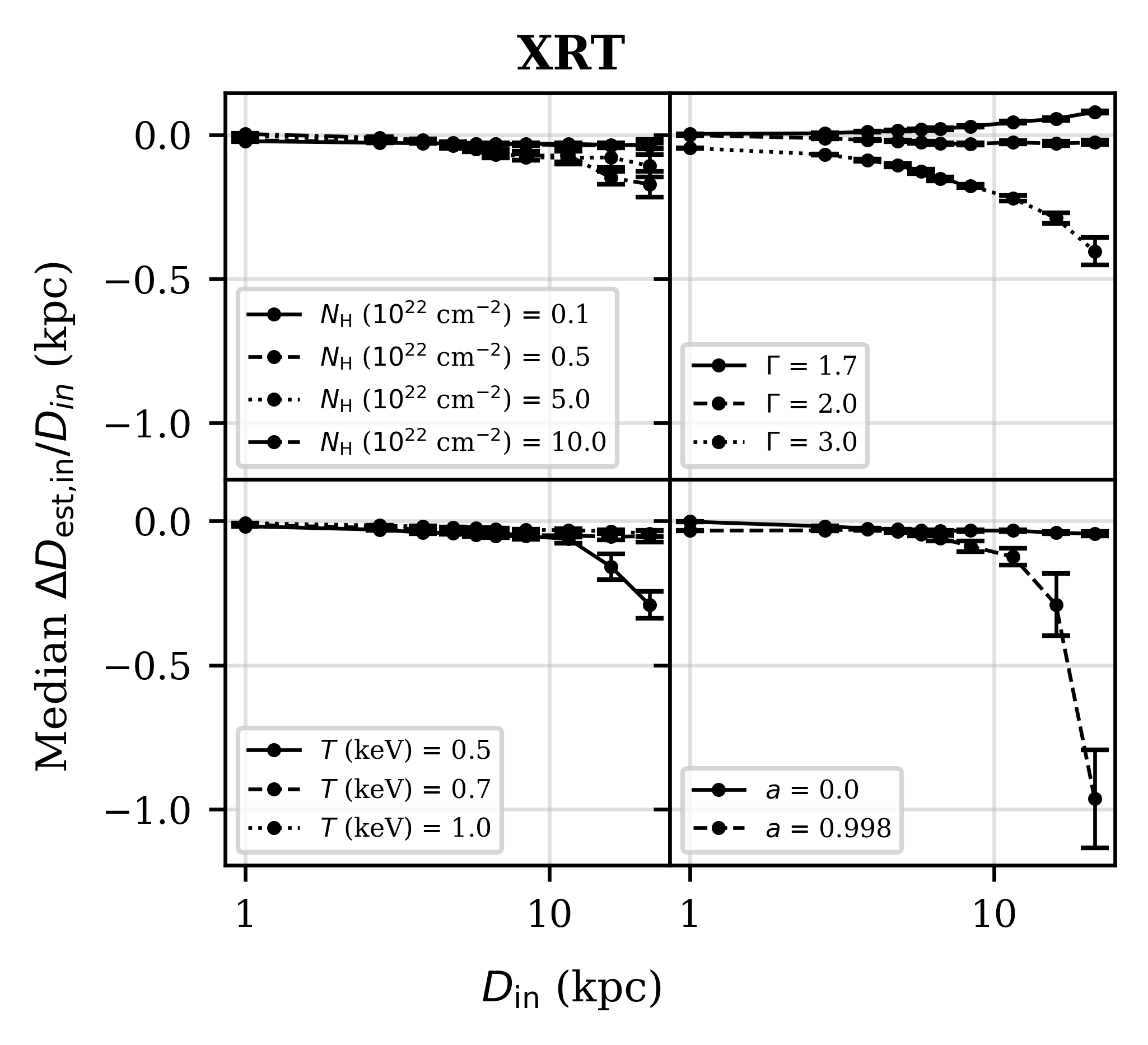}
        \label{fig:trendline1-xrt}
    \end{subfigure}
    \caption{The plot highlights the median fractional error between the estimated distance and the input distance ($\Delta D_{\text{est,in}}/D_{\text{in}}$) as a function of input distance ($D_{\text{in}}$). We present $\Delta D_{\text{est,in}}/D_{\text{in}}$ vs $D_{\text{in}}$ variation for the different values of: $N_{\text{H}}$ (top-left), $\Gamma$ (top-right), $T$ (bottom-left), $a$ (bottom-right). \textbf{Left.} Shows the results for \maxi/GSC simulations \textbf{Right.} Shows the results for \swift/XRT simulations. The error bars shown are the 68\% confidence errors on the median using 1000 bootstrap re-samplings.}
    \label{fig:combined_main_effects1}
\end{figure*}

We first look at the effects of every individual parameter on the estimated distance. We examine the effects by calculating the the fractional uncertainty:
\begin{equation}
    \frac{|\Delta D_{\text{est,in}}|}{D_{\text{in}}} =  \frac{|D_{\text{est}} - D_{\text{in}}|}{D_{\text{in}}}
    \label{eq:frac_uncert}
\end{equation}
for each parameter combination. Then, for each parameter (e.g., $\Gamma$ with input values $1,7,2.0, 3.0$), we calculate the median of the fractional uncertainty across the different input values of the parameter by ignoring the values of (marginalizing over) the other parameters. The resulting median $\Delta D_{\text{est,in}}/D_{\text{in}}$ thus provides an overall estimate of the error on the estimated distances for a given input parameter. In Figure~\ref{fig:combined_main_effects1}, we show the effects of $N_{\text{H}}$, $\Gamma$, $T$, and $a$ on the error for the estimated distances from \maxi/GSC (left plot) and \swift/XRT (right plot) simulations. The results from both instruments show a general trend of higher negative fractional error (estimated distance is less than the input distance) as the input distance increases. However, we note that there is one exception to this, the $\Gamma = 1.7$ simulations (top-right panel in the right plot of Figure~\ref{fig:combined_main_effects1}) in the \swift/XRT results, which shows an increase in the median positive fractional error. 

Comparing the results of \maxi/GSC with those of \swift/XRT, we observe that the changes in the fractional error with increasing input distance are steeper for \maxi/GSC. However, the $a=0.998$ case is an exception to this general trend, which shows a more sudden drop in the \swift/XRT estimated distance for the highest distance value (26\,kpc). Looking at the impact of the individual parameters on the accuracy of the estimated distances, we note that higher $N_{\text{H}}$ values (Figure~\ref{fig:combined_main_effects1}, top-left panel) affect \maxi/GSC more than \swift/XRT. We also note that the change of the accretion disc maximum temperature $T$ (Figure~\ref{fig:combined_main_effects1}, bottom-left panels) does not affect the median deviation of \swift/XRT results as much as their \maxi/GSC equivalents. 

\begin{figure*}
    \centering
    \begin{subfigure}[t]{0.49\textwidth}
        \includegraphics[width=\textwidth]{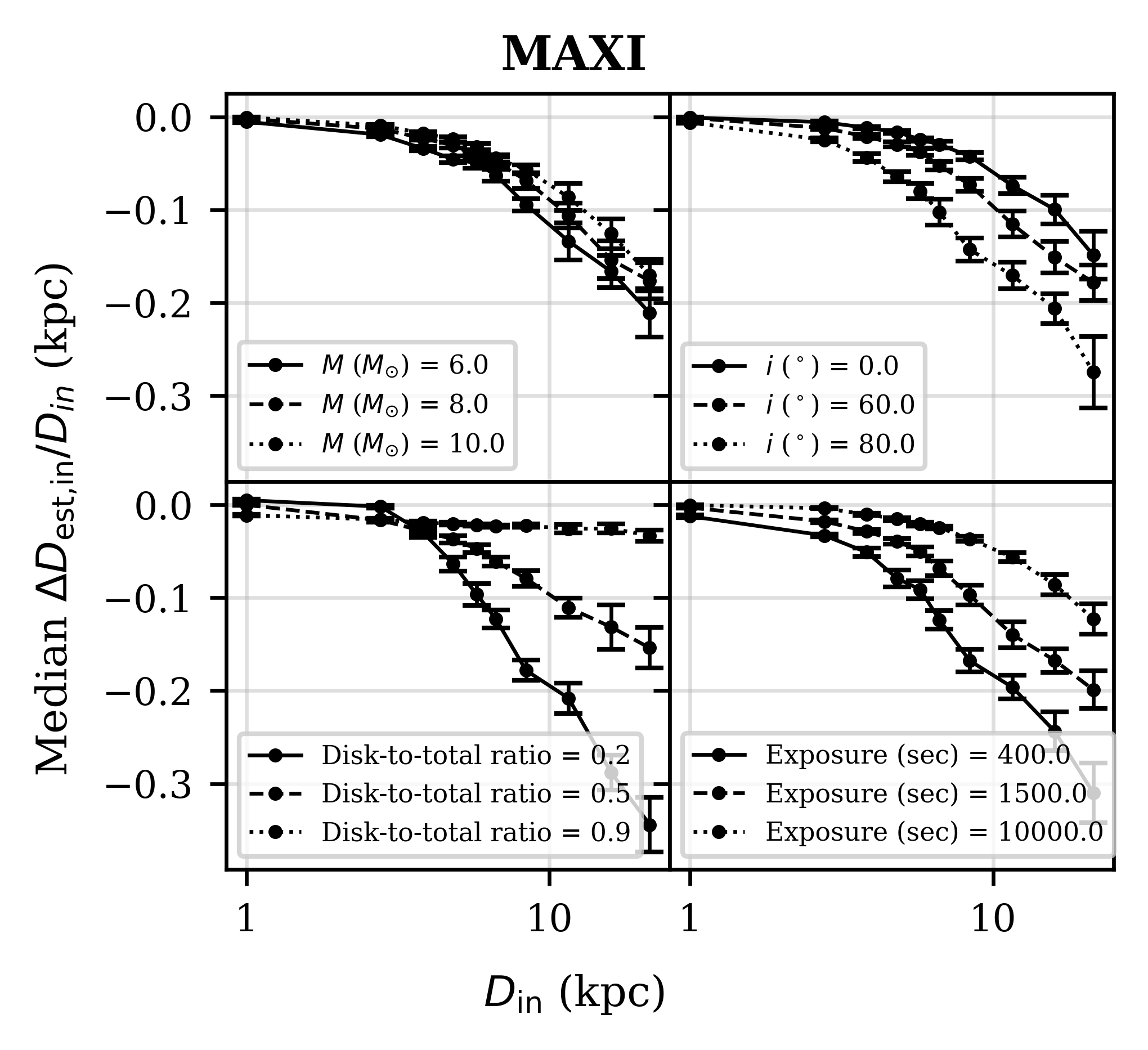}
        \label{fig:trendline2-maxi}
    \end{subfigure}
    \hfill
    \begin{subfigure}[t]{0.49\textwidth}
        \includegraphics[width=\textwidth]{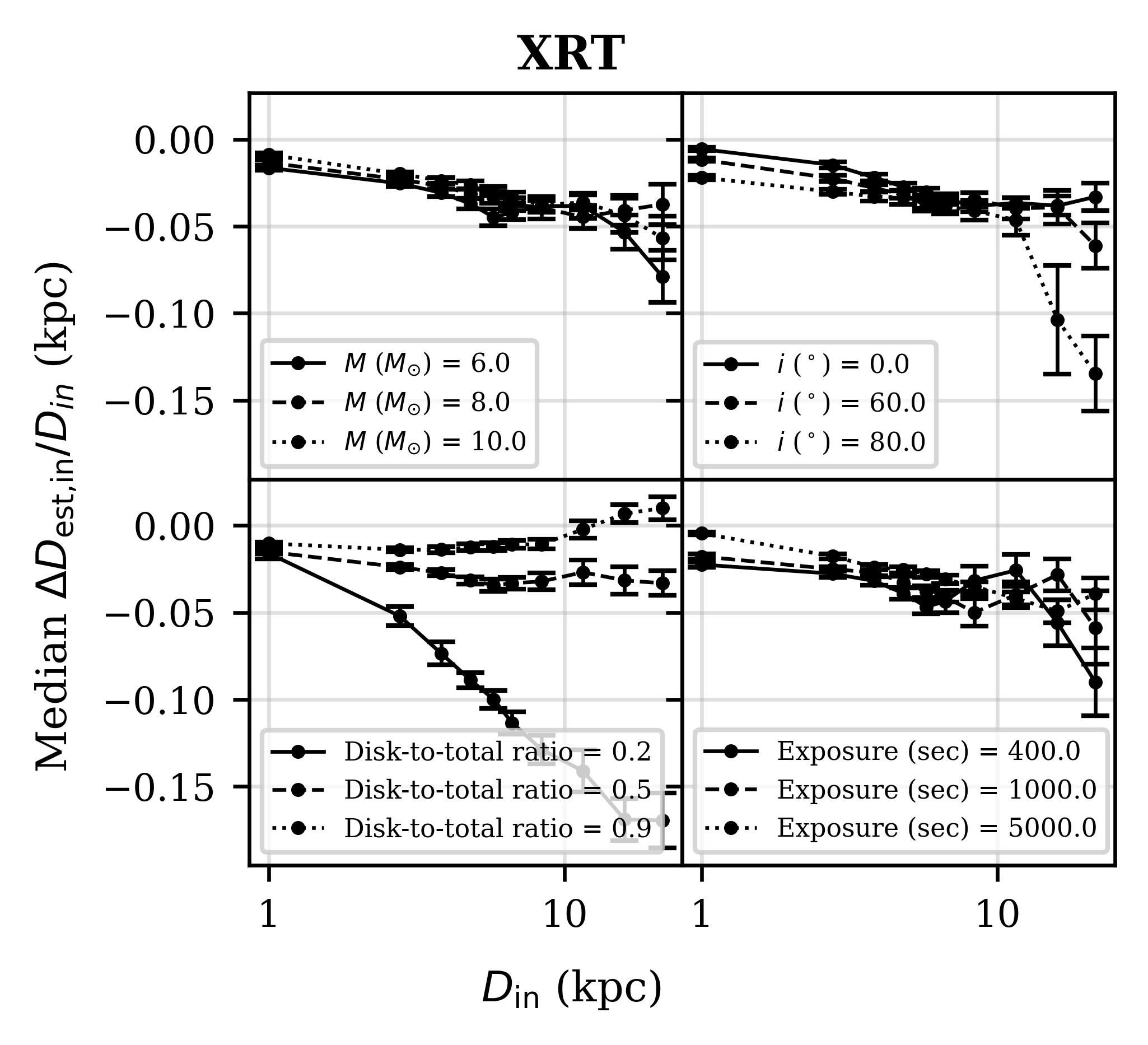}
        \label{fig:trendline2-xrt}
    \end{subfigure}
    \caption{Same as Figure~\ref{fig:combined_main_effects1} but for: $M$ (top-left), $i$ (top-right), disc-to-total ratio (bottom-left), Exposure time (bottom-right). \textbf{Left.} Shows the results for \maxi/GSC simulations \textbf{Right.} Shows the results for \swift/XRT simulations.}
    \label{fig:combined_main_effects2}
\end{figure*}

Similarly in Figure~\ref{fig:combined_main_effects2}, we show the effects of $M$, $i$, disc-to-total ratio, and exposure time. The general trend remains the same, where we observe an increasing negative fractional error with increasing distances, with the exception of the disc-to-total ratio 0.9 in the \swift/XRT simulations. The \maxi/GSC fractional error show stronger changes for all of these four parameters. We underscore that the exposure and black hole mass changes do not affect the median fractional error of the \swift/XRT simulations as much as they affect the \maxi/GSC ones. 

It is important to note here that the median fractional error results only show the general trends of the central tendency of the estimated distance. However, they do not capture the full picture of the estimated distance results. Thus, we also investigated the kernel density estimations (KDE) distributions of the deviations. To facilitate examining the differences more easily, we take the log-modulus transformation \citep{John1980} of the deviation $\Delta D_{\text{est,in}}$ (Equation~\ref{eqn:log-modulus}). This transformation preserves the sign of the deviation while taking the log of the deviation, which helps compress the plotting range and makes the plot easier to read. Figures~\ref{fig:a-dist}-\ref{fig:e-dist} in the appendix show the KDE distributions for all eight parameters. For brevity, we concentrate here on the most important findings that we can draw regarding the distributions from the KDEs. The \swift/XRT error distributions appear to be wider with less pronounced peaks and some bi-modalities than those from \maxi/GSC. However, the distributions central values appear closer to zero (i.e. the error in recovered distances is lower). In contrast, the \maxi/GSC distributions are narrower with well-defined single peaks tending toward larger negative deviations (input distance is more underestimated). A more comprehensive and detailed discussion about the error distributions can be found in appendix~\ref{app:distributions}. 

\subsubsection{Pairwise interaction effects between parameters}\label{sec:pairwise}

Having examined individual parameter effects, we now investigate how these parameters might interact to influence the uncertainty in our distance estimates. Specifically, we are interested in identifying significant pairwise interactions among the eight primary simulation parameters: $N_{\text{H}}$, $\Gamma$, $T$, $a$, $M$, $i$, disc-to-total ratio, and exposure time.

To achieve this, we model the logarithm of the absolute fractional uncertainty in the estimated distance, $Y = \log\left(\frac{|\Delta D_{\text{est,in}}|}{D_{\text{in}}}\right)$ with a linear regression model that tries to predict $Y$. The initial (full) model for $Y$ includes: (a) the true (input) distance ($D_{\text{in}}$) as a continuous variable; (b) the main effects for each of the eight simulation parameters; and (c) all $\binom{8}{2}=28$ pairwise interaction terms between these eight parameters. Given that the simulation parameters take discrete values, they are treated as categorical variables in the regression, implemented using dummy indicator variables (so they are either 0 or 1). A general representation of the full linear model is:
\begin{equation}
Y = \beta_0 + \beta_{D_{\text{in}}} D_{\text{in}} + \sum_{j=1}^{8} X_j(P_j) + \sum_{j=1}^{8}\sum_{k=j+1}^{8} I_{jk}(P_j, P_k)
\label{eq:linear_model}
\end{equation}
where 
\begin{itemize}
    \item $\beta_0$ is the intercept (baseline value of $Y$)
    \item $\beta_{D_{\text{in}}}$ accounts for the input distance
    \item $P_j$ and $P_k$ represent the simulation parameters
    \item $X_j(P_j)$ represents the main effect of parameter $P_j$ (involves multiple $\beta$ coefficients for its indicator variables)
    \item $I_{jk}(P_j, P_k)$ represents the interaction effect between parameter pair $P_j$ and $P_k$ (involving multiple $\beta$ coefficients)
\end{itemize}

To identify the most influential interaction terms from this comprehensive model, we utilise a backward stepwise selection procedure guided by the Bayesian Information Criterion \citep[BIC;][]{Schwarz1978}. This is a common technique for model selection that balances model fit with complexity, generally favouring more parsimonious models \citep[see, e.g.,][for discussions on model selection criteria]{Burnham2002, Liddle2007}. The backward stepwise process begins with the full model (all main effects, $D_{\text{in}}$, and all 28 pairwise interactions). Then, interaction terms are iteratively removed one at a time. At each step, the removal of each remaining interaction term is considered, and the term whose removal leads to the largest decrease (improvement) in the BIC value is eliminated. This process continues until no further removal of an interaction term can reduce the BIC. The interaction terms that remain in the final model are, therefore, the most important to obtain a good descriptor of the log fractional uncertainty variance.

Our modelling was performed on the logarithm of the absolute fractional uncertainty in the estimated distance as the transformed absolute fractional uncertainty satisfies the assumptions underlying the linear regression better, in particular, the normality of residuals \citep[e.g.,][]{Kutner2005}. Satisfying the assumptions strengthens the validity of our conclusions. For our modelling, we limited the simulation results to those where the fractional uncertainty (Equation~\ref{eq:frac_uncert}) is $\le 1$ (i.e., $\le 100\%$). This restriction is imposed because estimates with fractional uncertainties greater than 100\% are inherently unreliable and can introduce excessive noise, potentially masking or distorting the relative magnitudes of the interaction effects we aim to study. By excluding these data points, which would otherwise act as influential outliers and complicate the modelling process, we can build a more robust model that more clearly elucidates the importance of these interaction effects within a range of more realistic data.

The final model description obtained from our modeling for \maxi/GSC is the model where we have 13 significant interactions as listed in Table~\ref{tab:interactions}. In contrast, for the \swift/XRT simulations, the backward step process yielded the lowest BIC model with 11 interactions (Table~\ref{tab:interactions}).  

\begin{table}
\centering
\footnotesize
\setlength{\tabcolsep}{3pt}       
\renewcommand{\arraystretch}{0.9}  
\caption{Interactions included in the final models for \(\maxi/\mathrm{GSC}\) and \(\swift/\mathrm{XRT}\) simulations.}
\label{tab:interactions}
\begin{tabularx}{\columnwidth}{l*{4}{X}ZZ}
\toprule
 & \(a\) & \(i\) & \(T\) & \(\Gamma\) & Exposure & disc-to-total ratio \\
\midrule
\multicolumn{7}{l}{\(\maxi/\mathrm{GSC}\)} \\
\midrule
\(a\)            &               & \(\checkmark\) &              &              & \(\checkmark\) &              \\
\(T\)            & \(\checkmark\) &              &              &              & \(\checkmark\) & \(\checkmark\) \\
\(\Gamma\)       & \(\checkmark\) &              & \(\checkmark\) &              & \(\checkmark\) & \(\checkmark\) \\
\(N_{\text{H}}\) & \(\checkmark\) &              & \(\checkmark\) & \(\checkmark\) &              & \(\checkmark\) \\
\midrule
\multicolumn{7}{l}{\(\swift/\mathrm{XRT}\)} \\
\midrule
\(a\)            &               & \(\checkmark\) &              &              &              &              \\
\(T\)            & \(\checkmark\) &              &              &              &              & \(\checkmark\) \\
\(\Gamma\)       & \(\checkmark\) &              & \(\checkmark\) &              & \(\checkmark\) & \(\checkmark\) \\
\(N_{\text{H}}\) & \(\checkmark\) &              & \(\checkmark\) & \(\checkmark\) &             & \(\checkmark\) \\
\bottomrule
\end{tabularx}
\end{table}

\begin{figure} 
\includegraphics[width=0.95\linewidth]{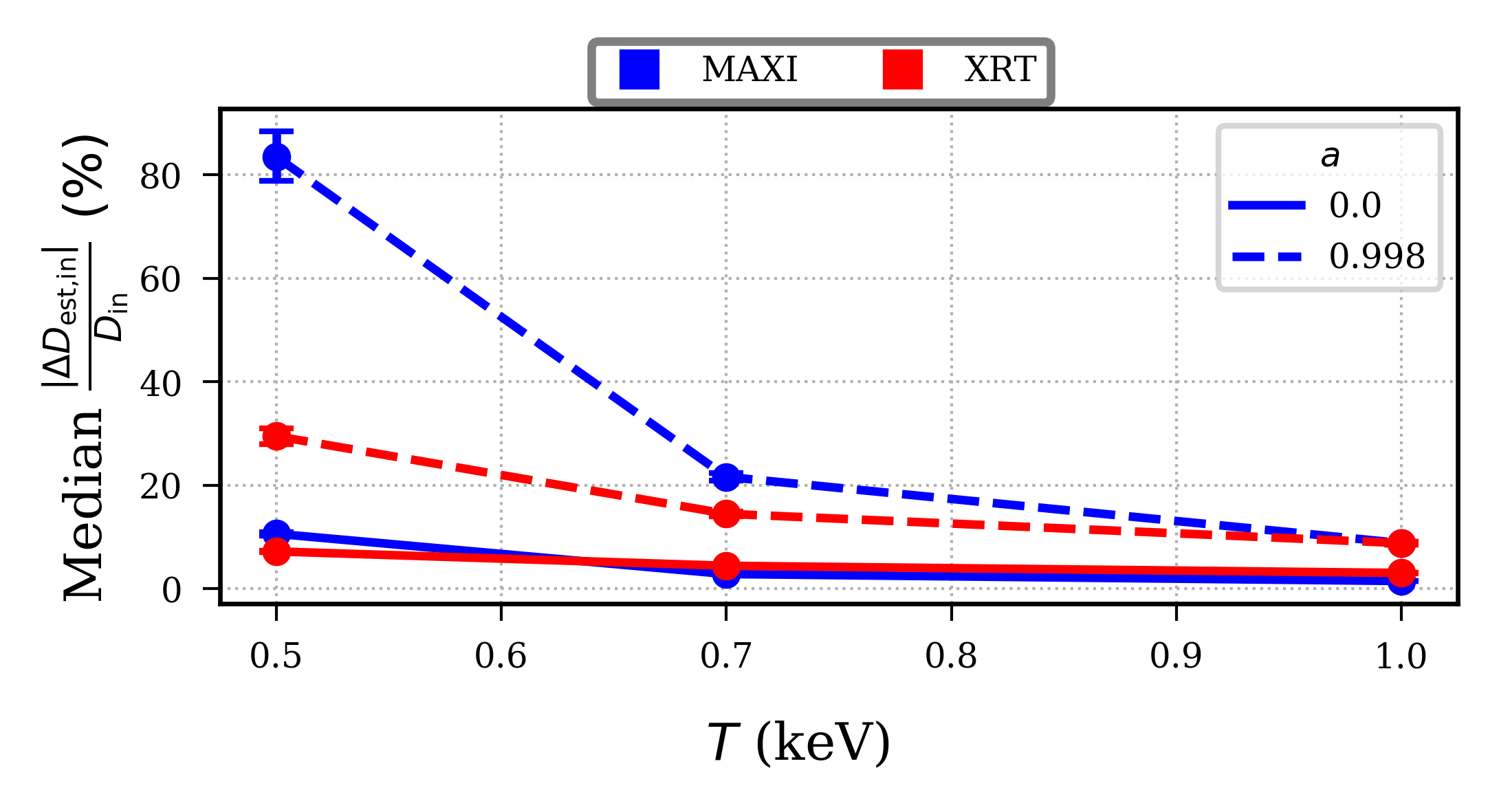} 
\caption{This figure shows the important pairwise interaction between the maximum disc temperature $T$ and the BH spin $a$. This was one of the interactions selected via the backward stepwise BIC process.} 
\label{fig:T-interactions-strong} 
\end{figure}

\begin{figure}
    \includegraphics[width=0.95\linewidth]{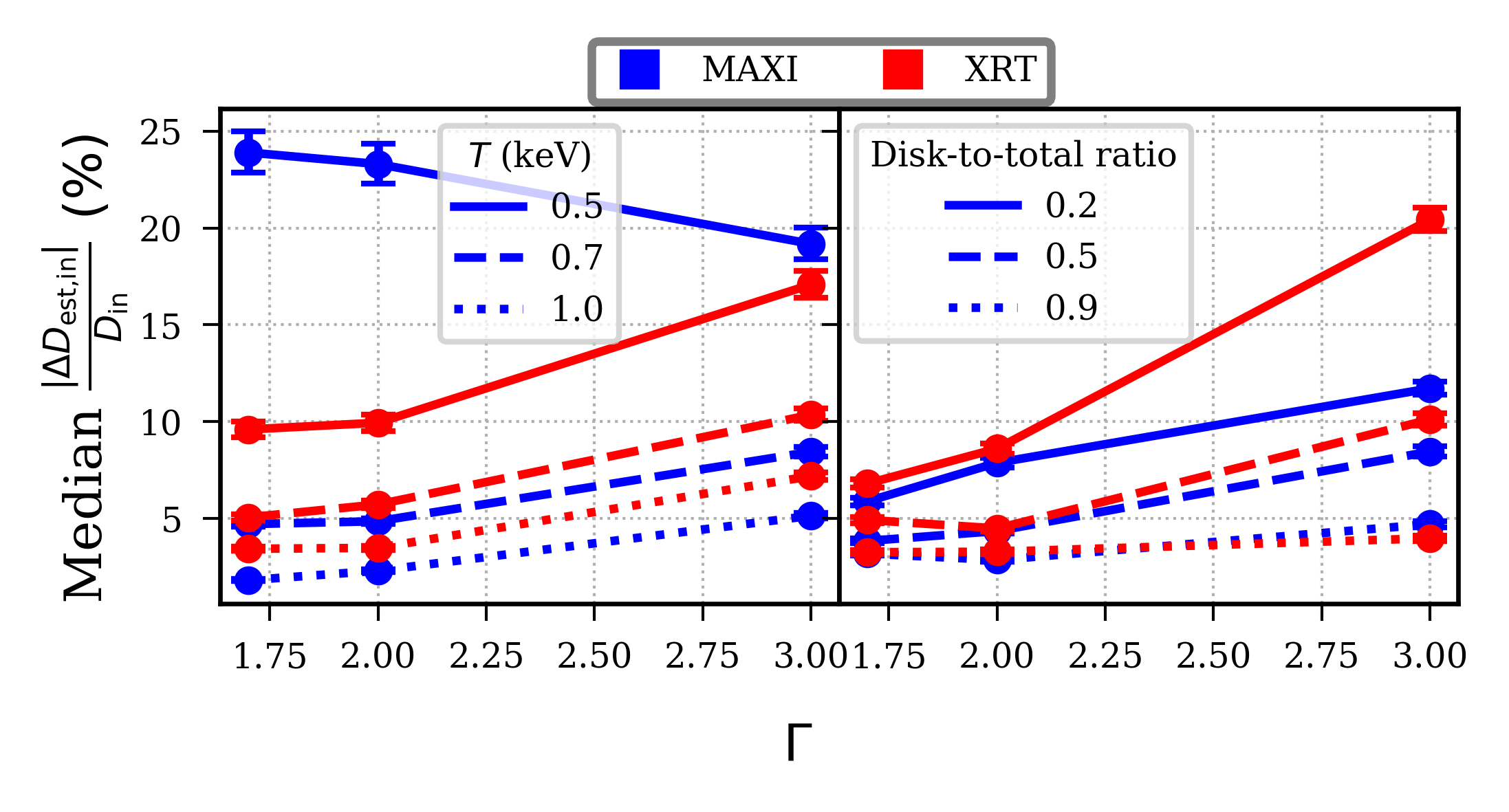}
    \caption{This figure shows the important pairwise interactions between the $\Gamma$ and other parameters. These interactions were selected via the backward stepwise BIC process.}
    \label{fig:g-interactions-strong}
\end{figure}

\begin{figure}
    \includegraphics[width=0.95\linewidth]{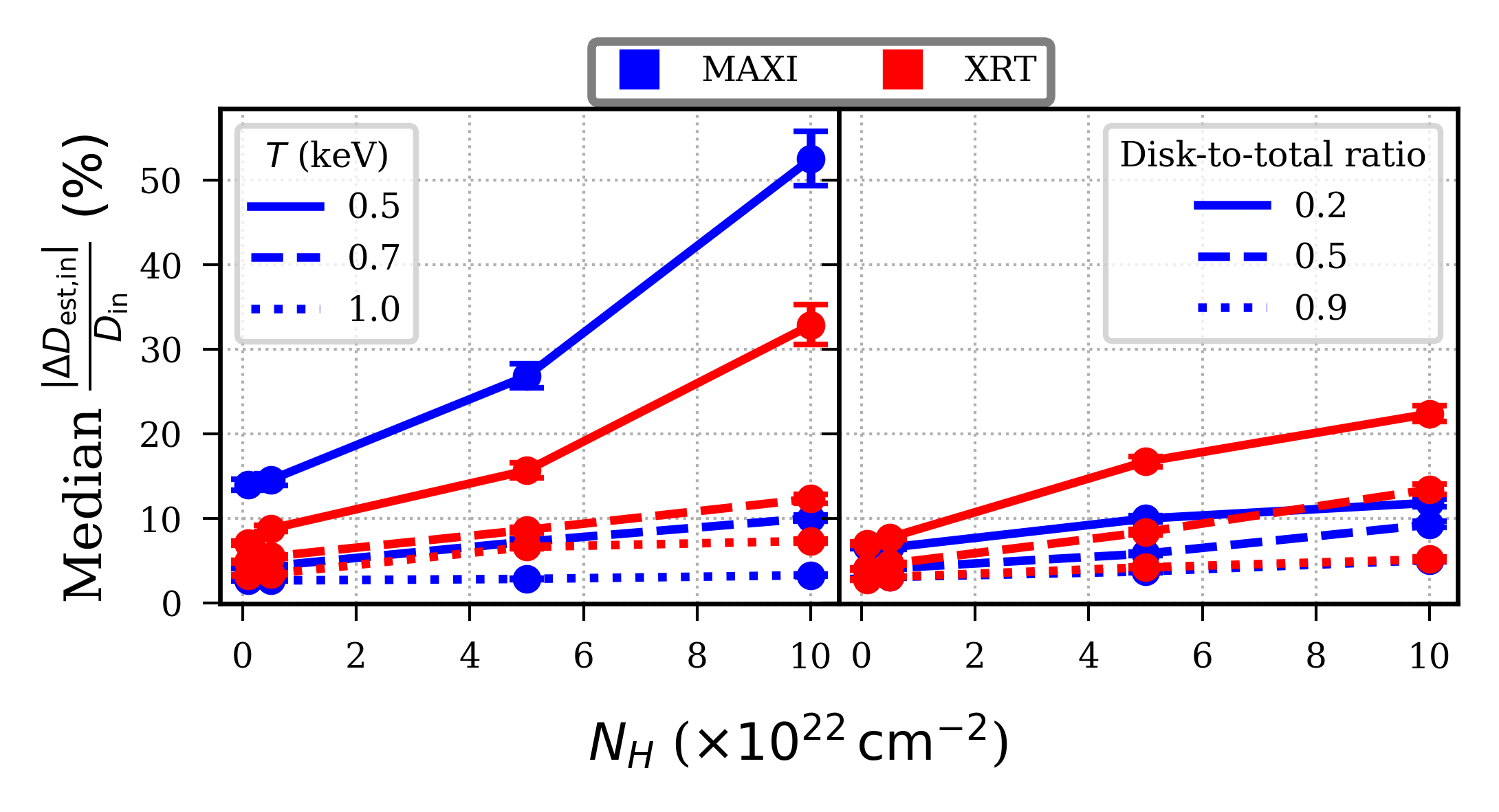}
    \caption{This figure shows the important pairwise interactions between $N_{\text{H}}$ and other parameters. These interactions were selected via the backward stepwise BIC process.}
    \label{fig:nH-interactions-strong}
\end{figure}

We observe that the interactions mostly follow the same patterns as in the individual parameter effects (see previous section). Thus, we succinctly show the most interesting interactions in this section while providing the rest of the BIC-selected interaction effects plots in appendix~\ref{app:interactions}. We construct the pair-wise interaction plot by using the fitted model equation, we calculate the predicted $Y$ for various combinations of some parameter $A$ and some parameter $B$. For clarity, we have transformed back the logged fractional uncertainty to fractional uncertainty (as a percentage) when we present the interaction plots (Figures~\ref{fig:T-interactions-strong}, \ref{fig:g-interactions-strong}, and \ref{fig:nH-interactions-strong}).

We first focus on a key interplay between the maximum disc temperature $T$ and the BH spin $a$, which offers further insight into the models behaviour. In Figure~\ref{fig:T-interactions-strong}, we plot the $T$:$a$ interaction. The figure shows that the increase in temperature when the BH is not spinning does not affect the relatively low median fractional uncertainty. In contrast, we observe that the increase from a disc temperature of 0.5 keV to 1 keV substantially improves the median fractional uncertainty of the high-spin simulations\footnote{This kind of behaviour, where the effect of one parameter changes when the other parameter value change is indicative of clear interaction between the two parameters that affect the response $Y$.}, especially for \maxi/GSC.  It is also to be noted from this figure and the rest of the interaction plots that the overall highest median fractional uncertainty from this model is obtained when $T=0.5$ keV and $a=0.998$ with \maxi/GSC having a median fractional uncertainty of $\sim80$\% at that combination, while \swift/XRT having a median fractional uncertainty $\sim30$\%. 

Another interesting interaction is the $\Gamma$:$T$ one, as shown in the left panel of Figure~\ref{fig:g-interactions-strong}. Here, the average fractional uncertainty generally increases with increasing $\Gamma$ and decreasing $T$. However, we observe an exception to this trend for \maxi/GSC results and the lowest temperature case, where at the highest $\Gamma$, we have a considerably lower fraction uncertainty. This ``anomalous'' decrease in the average fractional uncertainty for the highest $\Gamma$ and lowest $T$ is not present in the $\Gamma$:$T$ results for the \swift/XRT simulations, which shows a consistent trend. In the right-hand panel of Figure~\ref{fig:g-interactions-strong}, the $\Gamma$:disc-to-total ratio interaction is shown. The plot provides an intriguing result, where we see that for \swift/XRT, the combination of high disc-to-total flux ratio and high $\Gamma$ increases the fractional uncertainty substantially.

We observe another noteworthy finding in the $N_{\text{H}}$:$T$ interaction (left-hand panel of Figure~\ref{fig:nH-interactions-strong}). At disc temperatures of $T$ = 1 keV and 0.7 keV, the median fractional uncertainty changes minimally as the amount of absorption $N_{\text{H}}$ increases. Conversely, we observe that variations in the median fractional uncertainty exhibit more abrupt transitions at the lowest temperature. 

The last interesting interaction effect we mention is the $N_{\text{H}}$:disc-to-total ratio, where the fractional uncertainty of the \swift/XRT simulations shows a clear positive trend with increasing ratios and absorption. In contrast, the \maxi/GSC median fractional uncertainty shows a significantly weak positive trend.  

Figures~\ref{fig:T-interactions1}-\ref{fig:a-interactions2} in the Appendix show all remaining interactions and their effect on the median fractional uncertainty. 

As clear from the final BIC-selected models presented before, unlike the \maxi/GSC simulations, the $a$:Exposure and $T$:Exposure interactions were deemed unimportant by the BIC process for \swift/XRT. Generally, we notice that the median fractional uncertainties for \swift/XRT at the worst combinations are less than the \maxi/GSC results, except in a few cases, namely the $N_{\text{H}}$:disc-to-total ratio, $\Gamma$:disc-to-total ratio, and $N_{\text{H}}$:$\Gamma$, where we find that the magnitude of the median fractional uncertainties is higher than their \maxi/GSC counterparts.

\subsubsection{Overall distance recovery distributions}

\begin{figure}
    \centering
    \includegraphics[width=1\linewidth]{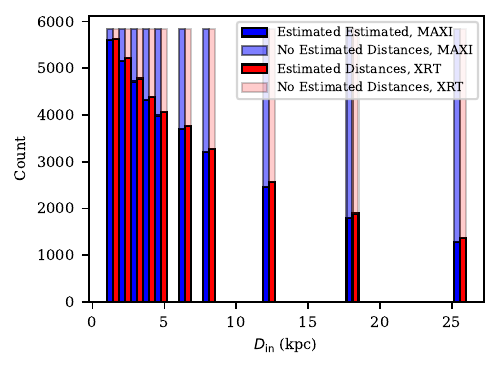}
    \caption{This plot shows a stacked histogram of the availability of distance estimations. This illustrates the overall ability of the instrument to provide a spectral fit that is sufficient to permit distance estimation. Note that for each $D_{\text{in}}$ we have a total of 4 (for each $N_{\text{H}}$ value)$\times 1458=5832$ simulations.}
    \label{fig:stacked_hist}
\end{figure}

To summarise our results, we show stacked histograms (Figure~\ref{fig:stacked_hist}) of the number of simulations per input distance with an estimated distance and the number of simulations input distances without an estimated distance. The lack of an estimated distance occurs due to the very low number of bins in the simulated spectrum, i.e. a low SNR in the spectrum. The number of available distance estimates appears to decay with increasing distance. While for \maxi/GSC, at the lowest distance, we have almost all of the simulated spectra (96\%) yielding a distance estimate. At the highest distance, only 1280 out of 5832 (22\%) simulations yield an estimate for the distance. In contrast, the decay in the number of estimable distances for \swift/XRT is slightly shallower. For example, looking at the number of the estimable distance at $D_{\text{in}} = 26$\,kpc, we observe a marginally larger number in the case of \swift/XRT compared to \maxi/GSC, with a count of 1364 as opposed to 1280 distances.

Restricting the estimated distances to only those with less than 50\% error from the input distances, we obtain empirical bias distributions of whether a fictitious source at a certain input distance could yield a distance estimate with an error of less than 50\% of the input distance. We convert these distributions to a probability density and model the density using possible candidate functions (Table~\ref{tab:pdfs}). We found that the function with the lowest mean squared error has power-law-like behaviour, in particular, the Pareto probability density function (PDF), which has the following form:

\begin{equation}
f(D_{\text{in}})= \,\frac{b}{(D_{\text{in}}-c)^{\,b+1}}
\label{eqn:best_pdf}
\end{equation}

\begin{figure}
    \centering
    \includegraphics[width=1\linewidth]{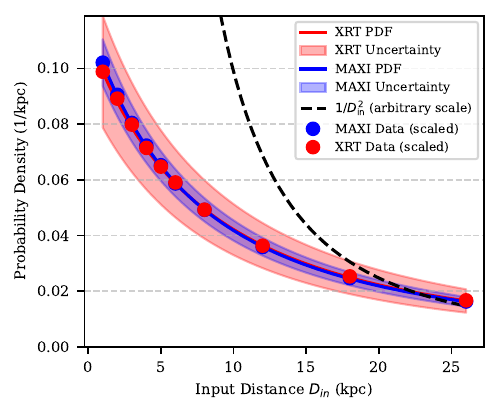}
    \caption{This figure shows the re-normalized fitted Pareto distributions to the density distribution of estimated distances that are within 50\% error of the input distance. The red curve is for \swift/XRT, and the blue curve is for \maxi/GSC. This distribution shows the overall ability of the instruments to provide accurate (with 50\% fractional error or not) distance estimations when using the soft-state method. The shaded regions correspond to the 1-$\sigma$ uncertainties from the curve fitting procedure. We also overlay the actual data points for each instrument and a scaled-$1/D^2$ relation. The fitted PDFs are very close to an inverse square law relation ($\propto 1/D^2$), at larger distances. Figure best viewed in colour.}
    \label{fig:PDFS}
\end{figure}

For \maxi/GSC the best-fit parameters were found to be: $b = 0.79\pm0.043$, $c=-12.88\pm0.45$. On the other hand, the best-fit parameters for \swift/XRT are: $b = 0.87\pm0.098$, $c=-14.62\pm1.075$. The normalised Pareto PDFs for both instruments are shown in Figure~\ref{fig:PDFS}. We note how close both distributions are to each other; we also observe that both distributions reduce, at large distances, to the usual inverse squared law relation $1/D^2$ expected for radiation. Since the PDFs from \maxi/GSC and \swift/XRT are almost identical, we take the average values of the best-fit values of both PDFs. We then use that in the next section (Section~\ref{sec:detailedobservationalview}) to correct the observed distance distribution obtained using the soft-state method in A24.

\section{The Corrected Observed Spatial Distribution}\label{sec:detailedobservationalview}

In the previous section, we derived an average bias PDF, $f_{\text{bias}}(D)$ (Equation~\ref{eqn:best_pdf}), which quantifies the likelihood of obtaining an accurate distance estimate (within 50\% error) at a given input heliocentric distance $D_{\text{in}}$ using the soft-state method from A24. We now leverage this $f_{\text{bias}}(D)$ to mitigate some of the systematic biases present in the original A24 distance estimates for our sample of Galactic BH-LMXBs.

The A24 study provided individual heliocentric distance probability density functions, $p_{\text{A24},s}(D_{\text{helio}})$, for each source $s$ in the sample. Our bias correction procedure is applied on a source-by-source basis to these individual $p_{\text{A24},s}(D_{\text{helio}})$. Specifically, for each source $s$, we define an unnormalized corrected distance PDF, $p'_{\text{corr},s}(D_{\text{helio}})$, by re-weighting its original A24 distance PDF by the inverse of our derived bias PDF:
\begin{equation}
p'_{\text{corr},s}(D_{\text{helio}}) = \frac{p_{\text{A24},s}(D_{\text{helio}})}{f_{\text{bias}}(D_{\text{helio}})}
\label{eq:bias_correction}
\end{equation}
where $D_{\text{helio}}$ represents the heliocentric distance variable. This re-weighting effectively up-weights distances where the A24 method might be less accurate (i.e., where $f_{\text{bias}}(D_{\text{helio}})$ is small), thereby attempting to compensate for the systematic under-representation of sources or distances that are harder to estimate accurately.

The resulting $p'_{\text{corr},s}(D_{\text{helio}})$ for each source is then re-normalized to ensure it is a proper probability density function that integrates to unity. The resulting $p_{\text{corr},s}(D_{\text{helio}})$ represents our best estimate of the bias-corrected heliocentric distance PDF for source $s$. From these individual corrected source PDFs, $p_{\text{corr},s}(D_{\text{helio}})$, we can derive the various quantities used in this section.

\begin{figure}
    \centering
    \includegraphics[width=1\linewidth]{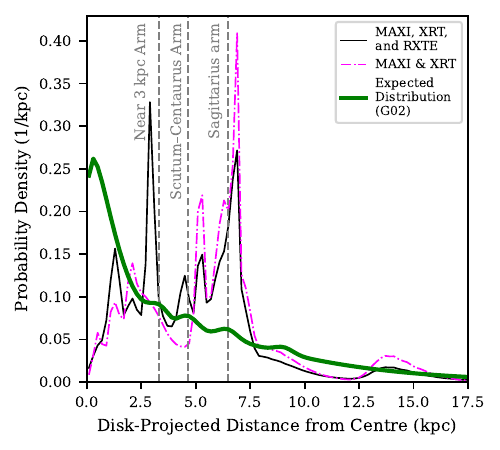}
    \caption{This plot shows the corrected and aggregated probability distribution of disc-projected Galactocentric distances using the \maxi~and \swift~sources (magenta line) and when including \rxte~sources (black line). The expected distribution from \citet{Grimm2002} (green line) is also shown. The vertical dashed lines mark the approximate locations of the near 3 kpc Arm, the Scutum–Centaurus Arm, and the Sagittarius Arm \citep{reid2019}. This figure is best viewed in colour.}
    \label{fig:gc_distance_distr}
\end{figure}

As A24's sample included sources observed by the Rossi X-ray Timing Explorer/Proportional Counter Array \citep[\rxte/PCA;][]{rxte} in addition to \maxi/GSC, and \swift/XRT. Therefore, we ideally also want to apply a bias correction to the \rxte/PCA distributions from A24. In order to avoid performing additional computationally expensive simulations for \rxte/PCA, we assume here that our average bias PDF obtained from the \maxi~and~\swift~simulations applies to \rxte/PCA. We expect that the realistic bias correction will be shallower since the response of \rxte/PCA is harder than \maxi/GSC.  Additionally, the effective area of only one\footnote{We only used Proportional Counter Unit (PCU) 2 in A24.} of the proportional photon counters (PCA) is around 1300 cm$^2$ which is an order of magnitude larger than \swift/XRT's ($\sim100$ cm$^2$) effective area \citep{rxte,xrt}. However, a quick investigation showed that applying a shallower profile to \rxte~distances will hardly affect our resulting PDF and subsequent conclusions.

In the A24 study, we presented a heliocentric distance distribution of all BH-LMXBs (Figure 3 in A24) by adding the best estimates of individual source distributions. In this current study, we improve this heliocentric distance distribution by utilising the corrected distribution and transforming the heliocentric coordinates to Galactocentric ones. In Figure~\ref{fig:gc_distance_distr}, we show the disc-projected Galactocentric distance probability distribution as observed by \maxi~and \swift~with the magenta curve. When we also extrapolate our bias corrector to the \rxte~sources, we show the corrected distribution containing all three inurements with the black line. The combined distribution with all three detectors shows peaks at $\sim 1.5$ kpc, 2.9\,kpc, followed by a dip in the 3-4\,kpc range. This is followed by a double-peaked structure around 5\,kpc. The density increases rapidly in the 6.5-7.5\,kpc region. Lastly, an elevated density is visible in the 12.5-15 kpc range. The distribution with only \maxi~and \swift~estimations displays the same major features but shows more a stair-like increase in the density in the 0-2.5\,kpc range and more pronounced dip in the 3-4 kpc range.

\begin{figure}
`	\includegraphics{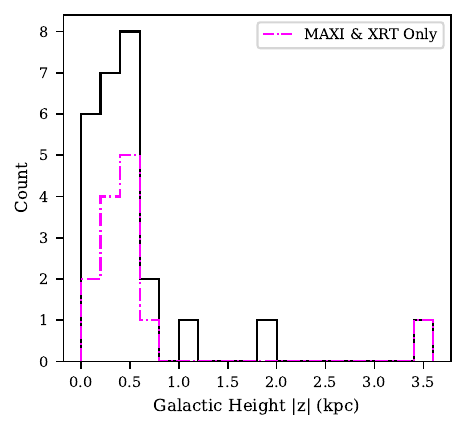}
    \caption{This plot shows a histogram (bin width = 0.2\,kpc) of the absolute galactic height $|z|$ for the sources in our sample using the best (with lowest 1-$\sigma$ errors) median distance estimates we obtained for each of them. The magenta line corresponds to \maxi~and \swift~sources only, and the black line corresponds to the results when including \rxte~source. The \texttt{astropy.coordinates} method was used to transform the (ICRS frame) RA, dec, and the distance into the galactocentric coordinates ($\rho,\phi,z$). The distance of the sun from the Galactic Center was set to 8.122 kpc, and the height of the Sun from the disc is 20.8 pc \citep{zsun}. This figure is best viewed in colour.} 
    \label{fig:height_distribution}
\end{figure}

Using the best (lowest 1$\sigma$ errors) bias-corrected median distance estimate we obtained for each of the sources as well as their right ascension and declination, we produced a histogram of the absolute galactic heights $|z|$ (Figure~\ref{fig:height_distribution}). Again, we separate the \maxi~and \swift~distribution from the one in which we include all detectors. Both histograms show that almost all sources in our sample have absolute galactic heights between 0.2 and 0.8 kpc, with only 3 sources having an absolute galactic height $\geq$ 1\,kpc.

\begin{figure}
	\includegraphics{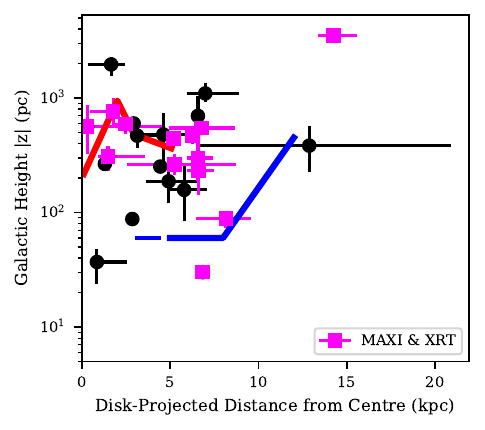}
    \caption{The plot shows the absolute galactic heights of our sample $|z|$ as a function of the disc-projected Galactocentric distance. This figure is adapted from the right panel of Fig. 1 in \citet{Jonker2021}. The red line approximates the shape and height of the Galactic Bulge \citep{portail2015,wegg2015}. While the blue line is 3 times the 20 pc scale height of massive stars in the region between 5 and 8 kpc from the galactic centre \citep{urquhart2014,reid2019}. Following \citet{Jonker2021}, we extend the line to the lower distance of 3\,kpc. While for larger distances between 8\,kpc and 12\,kpc, the scale height increases quickly to 450 pc. This figure is best viewed in colour.} 
    \label{fig:height_distribution_vs_rho}
\end{figure}

Furthermore, we adapt the right panel of Figure 1 in \citet{Jonker2021} and construct a similar graph of $|z|$ against the disc-projected Galactocentric distance (Figure~\ref{fig:height_distribution_vs_rho}).

The figure shows that the sample's Galactic heights are mostly contained in the region between 60 pc and 1000 pc. However, few sources appear to have higher or lower Galactic height by a factor of $\sim2-3$. 

Finally, for completeness, we provide an updated 2D map of the sources on the Milky Way plane in Figure~\ref{fig:2D-spatial_distribution}. We found that our general observations from A24 remain unchanged, where most sources are directed toward the Bulge, and there is a correlation with the spiral arms. 

\section{Discussion}\label{sec:discussion}

In this study, we performed extensive X-ray spectral simulations (Section~\ref{sec:simulation-study}) to evaluate the systematic biases in the A24 soft-state distance estimation method for BH-LMXBs. These simulations explored a wide parameter space, including intrinsic source properties (BH mass, spin, disc temperature, and disc-to-total flux ratio), and observational parameters (interstellar absorption, viewed inclination, and exposure time). We quantified the relation between accurate distance retrieval and input distance, which led to an empirical bias correction factor that, when applied to the A24 sample, yielded a de-biased Galactic BH-LMXB distribution (Section~\ref{sec:detailedobservationalview}). In this section, we first delve into the physical interpretation of the simulation trends and the overall robustness of the soft-state distance estimation method. We then compare our bias-corrected observed radial distribution with theoretical expectations. Further, we explore the role of interstellar absorption as an observational selection effect, potentially indicating a hidden BH-LMXB population, and conclude by examining the implications of our corrected Galactic height distribution for understanding natal kicks and BH formation channels.

\subsection{Physical interpretation of simulation results: expected trends}
\label{sec:physical_interpretation}

The simulation results presented in Sections~\ref{sec:sim_ind} and \ref{sec:pairwise} reveal distinct trends in how distance estimations deviate from the input distances based on the input parameters and their interplay. A critical question is how these dependencies and interactions can be physically understood, given the underlying A24 distance estimation method, the assumed X-ray spectral model, and the characteristics of the simulated instruments (\maxi/GSC and \swift/XRT). Overall, we find that the observed trends are largely consistent with established physical principles of X-ray emission from accretion discs, X-ray absorption, and detection limitations. The A24 method's reliance on accurately modeling the thermal disc component (\texttt{ezdiskbb}), particularly its normalisation ($N_{\text{ezdiskbb}}$ in Equation~\ref{eq:norm_ezdiscbb_corrections}), is central to understanding the observed effects.

For instance, higher Galactic column densities lead to an underestimation of the distance (Figure~\ref{fig:combined_main_effects1}) because the absorption of soft X-rays appears to weaken the disc component, which typically peaks at energies $\lesssim 2$~keV. This effect is more pronounced for distant sources, or instruments which are less sensitive at low energies, as absorption can render the disc too faint or its shape too ill-defined for a good characterisation of the disc spectrum. Similarly, lower maximum disc temperatures shift the disc's emission peak to softer energies. This makes the emission more susceptible to interstellar absorption and can move a significant portion of the disc flux out of the primary sensitivity range of the instruments, which in turn results in a poorly constrained $N_{\text{ezdiskbb}}$ and, consequently, an underestimation of the distance. Conversely, higher temperatures ($T \gtrsim 0.7-1.0$~keV) produce brighter disc spectra that peak at higher energies, making them more distinguishable and robustly constrained against absorption and instrumental cut-offs, thus improving $N_{\text{ezdiskbb}}$ determination.

The BH spin $a$ also plays a crucial role as it changes the inner disc radius. The disc-to-total flux ratio directly determines the disc component's prominence. A low ratio (e.g., 0.2, Figure~\ref{fig:combined_main_effects2}) implies that the spectrum is dominated by the power-law component, making the weaker disc emission difficult to isolate and model accurately for $N_{\text{ezdiskbb}}$ determination. A high ratio (e.g., 0.9), on the other hand, ensures the disc is dominant, allowing its parameters, including $N_{\text{ezdiskbb}}$, to be more easily and accurately constrained, leading to more reliable distance estimates.

The power law photon index $\Gamma$ influences spectral decoupling. If the disc is not strongly dominant, the power-law might contribute significantly, even in softer bands, potentially biasing the disc fit. A very soft power law ($\Gamma \sim 3.0$) can also be spectrally similar to a cool disc component, leading to degeneracies if the disc is weak or cool, thereby affecting $N_{\text{ezdiskbb}}$ (Figure~\ref{fig:combined_main_effects1}). The unusual positive deviation observed for $\Gamma=1.7$ with \swift/XRT might indicate scenarios where fitting a hard power law in a disc-dominated spectrum forces an overestimation of any residual soft component if the decoupling of the components is imperfect. For a given spin, a larger BH mass ($M$) implies a larger $r_{\text{in}}$ (as $r_{\text{in}} \propto M$), resulting in a larger emitting area and typically a higher intrinsic disc luminosity and normalisation. Such a disc is generally easier to detect and model accurately (Figure~\ref{fig:combined_main_effects2}), leading to potentially smaller distance deviations, particularly for instruments like \maxi/GSC where sensitivity is a stronger limiting factor. Higher inclinations reduce the projected emitting area of the disc (by a factor of $\cos i$) and involve relativistic effects (accounted for in Equation~\ref{eq:norm_ezdiscbb_corrections}). The net result is a lower observed flux for a given intrinsic luminosity and distance, making the disc parameters harder to constrain accurately. Lastly, longer exposure times yield higher SNR spectra (Figure~\ref{fig:combined_main_effects2}), generally improving the accuracy of spectral fitting, and thus more reliable distance estimates. It is noteworthy that \swift/XRT's higher sensitivity allows it to maintain good SNR spectra even at intermediate distances and with moderate exposures (Figure~\ref{fig:combined_main_effects2}), as expected when compared to \maxi/GSC.

The significant pairwise interactions identified (Section~\ref{sec:pairwise}) seem to be broadly consistent with physical expectations, typically highlighting scenarios where combined parameter effects exacerbate challenges in the spectral fitting process. For instance, the interaction between interstellar absorption and disc temperature ($N_{\text{H}}$:$T$, Figure~\ref{fig:nH-interactions-strong} left panel) is critical. When $T$ is high ($\ge 0.7$~keV), the disc emits substantially at energies less affected by $N_{\text{H}}$. However, for cool discs ($T \le 0.5$~keV), which peak at soft energies, increasing $N_{\text{H}}$ severely attenuates the observable disc flux, making $N_{\text{ezdiskbb}}$ determination highly unreliable. This explains the sharp increase in fractional uncertainty for the $T=0.5$~keV case as $N_{\text{H}}$ rises. Interactions involving the photon index with disc temperature ($\Gamma$:$T$) or with the disc-to-total ratio ($\Gamma$:disc-to-total ratio, Figure~\ref{fig:g-interactions-strong}) reflect the challenges of spectral decoupling. When the disc component is intrinsically weak (due to low $T$ or a low disc-to-total ratio), separating it from either a hard power law or a very soft power-law becomes problematic, leading to increased uncertainties in $N_{\text{ezdiskbb}}$. The specific trend for \swift/XRT where high $\Gamma$ combined with a high disc-to-total ratio still shows increasing uncertainty (Figure~\ref{fig:g-interactions-strong} right panel) is intriguing and may suggest that even a minor but very steep power law component can complicate the baseline determination for a dominant soft disc, or affect the broader continuum fitting that $N_{\text{ezdiskbb}}$ depends on.

In summary, the simulation results align with physical and instrumental expectations. The observed deviations and uncertainties primarily stem from predictable physical processes that either reduce the quality of the spectral data or complicate the isolation and modelling of the crucial thermal disc component.

\subsection{Robustness of the soft-state distance estimations}\label{sec:robustness-discussion}
One of the outcomes of our simulation study is that we now have a more precise understanding of the robustness of the soft-state distance estimation method developed in A24. The results of the simulations show that, in general, the method is quite robust against any variations in the eight simulation parameters ($N_{\text{H}}$, $\Gamma$, $T$, $a$, $M$, $i$, disc-to-total ratio, exposure time) at low distances, namely, the 1--3\,kpc range. Beyond that, the robustness of the method becomes highly dependent on the specific combination of these parameters, which, given the sheer number of their possible combinations, is impossible to investigate completely. Nevertheless, we can still discuss the general findings that we observed.  

One of the findings was that the method had increased robustness against variations in other parameters when the disc-to-total flux was 0.9 (see Figures~\ref{fig:combined_main_effects2},~\ref{fig:r-dist}). This finding is unsurprising, as our distance estimation relies on the disc parameters, but reassuring since typically, in the soft-state, the disc flux ratios are expected to be $>0.75$ \citep{Remillard2006}.

\begin{figure}
    \centering
    \includegraphics[width=1\linewidth]{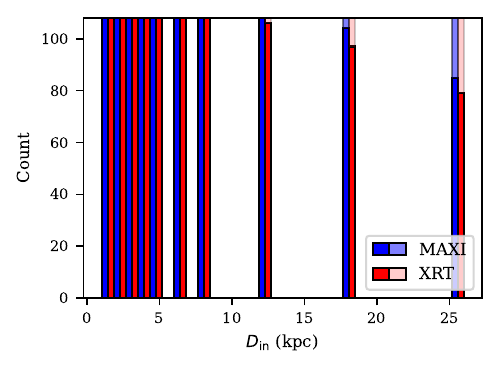}
    \caption{This plot shows a stacked histogram of the availability of distance estimations that are within 10 \% of the input distance for the subset of simulations with; disc-to-total flux ratio equal to 0.9, disc maximum temperature of 1 keV, non-spinning BHs, and the maximum exposure for each instrument (10 ks for \maxi/GSC and 5 ks for \swift/XRT). Note that for each $D_{\text{in}}$ we have a total of 108 other parameter combinations.}
    \label{fig:stacked_hist_r0.9_T1_eMAX_nospin}
\end{figure}

Furthermore, if we only look at the subset of data with a disc-to-total flux ratio equal to 0.9, we can still identify some influential values of other parameters that affect the deviations from the input distance by a substantial amount. Therefore, we can again look at the ``most'' robust parameter value and select the subset of the simulations with that value. Of course, we can follow this process iteratively to hone in on the most robust combinations. In particular, our simulation data show that if we follow that kind of iterative procedure, we can find that the following subset; disc-to-total flux ratio equal to 0.9, disc maximum temperature of 1 keV, non-spinning BHs, and the maximum exposure for each instrument (10 ks for \maxi/GSC and 5 ks for \swift/XRT), have very good recovery of input distances (fractional error is $\leq$ 0.1) for all input distances up to 12 kpc. This is independent of the values of the other parameters and the instrument (Figure~\ref{fig:stacked_hist_r0.9_T1_eMAX_nospin}).  

\begin{figure}
    \centering
    \includegraphics[width=1\linewidth]{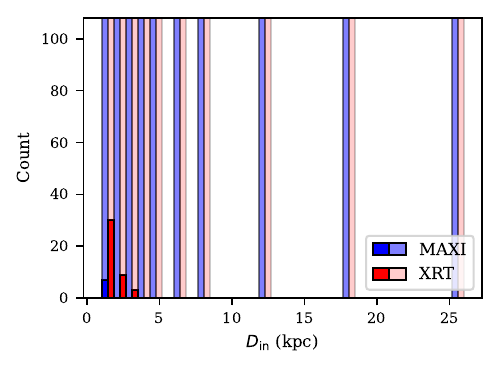}
    \caption{Same as Figure~\ref{fig:stacked_hist_r0.9_T1_eMAX_nospin} but for the subset of simulations with; disc-to-total flux ratio equal to 0.2, disc maximum temperature of 0.5 keV, maximal-spinning BHs, and the minimum exposure of 0.4 ks.}
    \label{fig:stacked_hist_r0.2_T0.5_eMIN_maxspin}
\end{figure}

We can also similarly investigate the opposite question. What are the most alarming combinations? A hint was already given when we found that in the results of the pairwise interaction effects, the simulations of $T=0.5$ keV and $a = 0.998$ gave the highest median fractional uncertainty. Indeed, we find that, simply by picking the opposite extremum, namely, the subset with disc-to-total flux ratio equal to 0.2, disc maximum temperature of 0.5 keV, maximal-spinning BHs, and minimum exposure of 400\,s. A good recovery of the input distances is almost non-existent, especially for the \maxi/GSC simulations (Figure~\ref{fig:stacked_hist_r0.2_T0.5_eMIN_maxspin}).

These results are somewhat intuitive if we consider that accurate estimates are only obtained when the input normalisation of the \texttt{ezdiskbb} model is recovered sufficiently well from the simulated spectra, that is subject to a certain SNR binning, background subtraction, and the instrument's fixed effective energy range. For instance, it is simple to show that by lowering the disc-to-total flux ratio and observation exposure time, the SNR binned spectra become a much lower-quality representation of the underlying input spectrum. Furthermore, it can also be shown that a lower maximum disc temperature shifts the \texttt{ezdiskbb} curve out of the detector energy range. Thus, the observed portion of the disc component becomes smaller relative to the power law, which at low normalisation (and lower flux) can cause the fit to confuse between the two. Lastly, although less obvious, we argue that the maximal spin case causes the normalisation to be lower compared to the zero spin case when the other parameters in Equation~\ref{eq:norm_ezdiscbb_corrections} are kept constant and given the inclination range, we considered\footnote{We did not explicitly show how does the spin-dependant factors in Equation~\ref{eq:norm_ezdiscbb_corrections} change since this is beyond the scope of this paper. But we refer the reader to \citet{zhang1997} and \citet{salvesen2021} for a more rigorous investigation}. This decrease in normalisation again causes the simulated spectrum to be more affected by the subsequent background ``filtering'' process. 

In section~\ref{sec:detailedobservationalview}, we utilised the resulting probability distribution (Figure~\ref{fig:PDFS}) of the detector ``bias'' with distance to calibrate the method and mitigate some of the effects resulting from the aforementioned spectral changes. The negative offset $c$ in the Pareto function's denominator (Equation~\ref{eqn:best_pdf}), $(D_{\text{in}}-c)^{b+1}$ (effectively $(D_{\text{in}}+|c|)^{b+1}$), results in a gentler decrease in accuracy probability at smaller input distances ($D_{\text{in}}$ not significantly larger than $|c|$), compared to a simple $D_{\text{in}}^{-(b+1)}$ fall-off. This flattening is physically reasonable: at closer distances, the SNR is generally high across a wide array of simulated spectral parameter combinations. Consequently, the probability of an accurate distance estimate is already high and exhibits less steep dependence on minor variations in $D_{\text{in}}$. In this high-SNR regime, factors other than pure photon statistics, such as inherent complexities in spectral components decoupling even at good SNR, may become more influential in defining the upper bounds of accuracy probability. The $1/D_{\text{in}}^2$-like dependence at larger distances, we think, is fundamentally rooted in the geometric dilution of flux from the source, which decreases as $1/D_{\text{in}}^2$. For relatively constant background noise, the SNR for the thermal disc component vital to the A24 distance estimation method, will exhibit a strong decrease with distance, scaling similarly to the flux ( $\propto 1/D^2$).

Based on the results of the simulations, we confirm that the soft-state distance estimate procedure developed in the A24 framework is considerably more robust when a good fit to the spectrum returns a high disc-to-total flux ratio $\sim0.9$ and a maximum disc temperature $\gtrsim 1$ keV.  

\subsection{How does our bias-corrected observed disc radial distribution compare to the expected distribution?}

In Figure~\ref{fig:gc_distance_distr}, we overlaid the theoretically expected distribution as modelled by \citet[][hereafter, G02]{Grimm2002} for the Galactic LMXB population. The G02 distribution assumes that the LMXB distribution (both BH and neutron star) should follow the empirically constrained stellar distribution models of the Milky Way \citep{Grimm2002}. Therefore, they use the same three components (bulge, disc, and spheroid) to model the standard stellar distribution \citep{Bahcall1980,Dehnen1998}. However, they modify it to better fit the observed LMXB population by changing the vertical scale height of the disc component and re-weighing the relative ratio between the disc, and the spheroid masses \citep{Grimm2002}. It is important to note that this expected distribution is obtained by analysing both BH and neutron stars, but our sample does not include neutron star LMXB. However, we believe that this should still be a highly insightful comparison since there would likely be a correlation between the BH-exclusive distribution and the neutron star one. This is because of two main reasons: their massive progenitors should be correlated \citep{fryer2001}, and a BH-LMXB can form by mass fallback/accretion to an initial neutron star in an LMXB \citep[e.g.][]{Wong2014}. 

We follow the same parametrisation they use in their Equations (4-6, and 10) and take the value of most parameters from Table 4 in their papers. However, we use the mass density normalisations from \cite{Atri2019}, since they were not explicitly provided in G02's Table 4. Moreover, to approximately reproduce the G02 distribution and compare it to our results, we convert the mass density to 1D probability by first assuming that the number density follows the same relation, and we then multiply by the Jacobian to transform the volume density to a radial density. Furthermore, we follow their model of the spiral structure, where they increase the disc density by a maximum of 20\% utilising an angle-dependent Gaussian centred on the spiral arms locations (equation 10 in G02). We take the locations of the spiral arms model to be equally spaced (initially) from the Galactic centre. Lastly, we also included the approximate locations of the Near 3-kpc, Scutum Centaurus, and Sagittarius arms based on the most recent understanding of our Galaxy structure \citep{reid2019}. 

The figure shows good agreement with the G02 model, especially when considering the locations of the spiral structure and the decay with distance from the Galactic Centre. Nevertheless, there is a clear paucity in the near-centre sources. Based on the arguments by \citep{Jonker2021}, and our simulations (see section~\ref{sec:hidden_population} and Figure~\ref{fig:nH-map_non_estimable}), we think that this discrepancy might be an observational selection effect associated with the high interstellar absorption along the line of sight at low galactic heights. Alternatively, although less likely, this discrepancy could also be due to a true reduction in the number of BH-LMXBs in the region very close to the Galactic Centre. This might be because the star population very close to the centre is old, and thus the probability of having massive primaries (i.e. BH progenitors) is lessened compared to the disc, which has more active regions and young populations. 

\subsection{Interstellar absorption observational effect: a hidden population?}\label{sec:hidden_population}

The interstellar absorption along the light-of-sight to an X-ray binary correlates with its location relative to the galactic disc. It is well established that generally the further from the galactic disc a source is observed, the lower its interstellar absorption/extinction will be \citep{Kalberla2005,HI4PI2016,Juvela2016}. Thus, the interstellar absorption can be considered as a proxy for the galactic height from the disc plane of a BH-LMXB. Furthermore, natal kicks during BH formation mean that even if a BH originated in the disc, it may end up at high galactic heights above and below the plane \citep{Jonker2021}. The determination of the galactic heights and possible associated selection effects is therefore important to constrain the natal kick distribution and, subsequently, BH formation theories.

\begin{figure}
    \centering
    \includegraphics[width=1\linewidth]{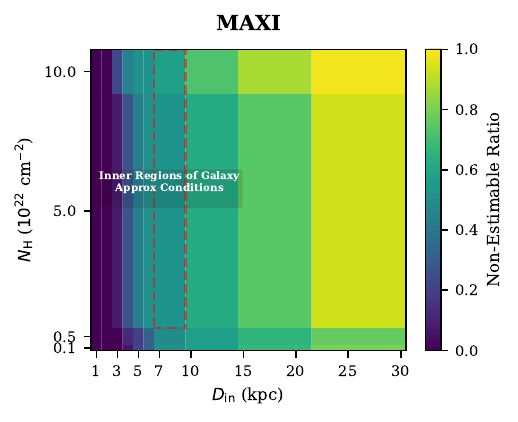}
    \caption{This 2D heatmap shows a distance mapping of the effect of $N_{\text{H}}$ on the ability of the method to provide an estimated distance for the soft-state simulated spectrum and \maxi/GSC data. This correlates with the ability of the instrument to ``detect'' sources. The non-estimable ratio is calculated by taking the count of the combinations that had no distance estimates over the total combinations (54) at each $N_{\text{H}}-D_{\text{in}}$ grid point. The semi-transparent red dashed rectangle highlights an approximate region corresponding to conditions typical for sources in the inner region of the Galaxy ( $D_{\text{in}} \approx 6.5-9.5$~kpc and $N_{\text{H}} \gtrsim 10^{22.5}$~cm$^{-2}$)}.
    \label{fig:nH-map_non_estimable}
\end{figure}

The combination of interstellar extinction, distance from the source, and peak X-ray flux during the outburst could all conspire to produce a possible selection effect on the observed BH-LMXB population and their Galactic heights. 

Our simulation study, which considers a combination of interstellar absorption, distance to the source, intrinsic properties of the source, and inclination, provides evidence that interstellar absorption plays an important role in detectability. 

To show this, we used only the \maxi/GSC simulations since it is an actual all-sky monitoring telescope used to detect BH-LMXBs in contrast to \swift/XRT, which is usually used for follow-up observations. Furthermore, we limit the simulations to the ones that are close to the spectra typically observed in the soft intermediate state (i.e. the intermediate state before the soft state) in BH-LMXBs so that the source is, in principle, at or near the brightest phase during its outburst. In particular, we chose the \maxi/GSC simulations with a disc-to-total ratio of 0.5 and with a maximum disc temperature of 0.7 keV. Lastly, we use the 1500\,s exposure time, which is very close to the average daily exposure time ($\sim$ 1400\,s) for a typical single X-ray source in \maxi/GSC observations \citep{Sugizaki2011}. Since distance to the source and the magnitude of $N_{\text{H}}$ are coupled in their effect on detectability, we plotted a 2D heatmap (Figure~\ref{fig:nH-map_non_estimable}) that represents the degree of ``undetectability'' at each $N_{\text{H}}$-$D_{\text{in}}$ simulation grid point. The coupling effect is evident when one notices that at a very low distance of 1-2 kpc, there is virtually no effect of $N_{\text{H}}$ on the number of detectable sources. Once the source is put at a distance larger than 3\,kpc, we find that the number of detectable sources in the range of 4-6 kpc are greatly suppressed at high interstellar extinction compared to low interstellar extinction. At input distances $\geq 8$ kpc, the increase in distance is the dominant effect, but interstellar extinction still plays some role in detectability. A simple comparison in the distance range $3-12$\,kpc, shows that the number of undetected sources almost doubles (161 instead of 87) at $N_{\text{H}} = 10^{23}$ \pcm~compared to $N_{\text{H}} = 10^{21}$ \pcm. 

This finding thus suggests that interstellar absorption is one of the main observational selection factors that can cause the low number of sources observed at very low heights ($|z| \lesssim 0.02$ kpc) relative to the Galactic disc plane due to the high interstellar absorption in that region \citep{Willingale2013,HI4PI2016}. Figure~\ref{fig:nH-map_non_estimable} visually underscores this: the overlaid region, indicative of typical distances and high absorption columns expected for sources towards the inner Galaxy (e.g., $D_{\text{in}} \approx 6.5-9.5$~kpc and $N_{\text{H}} \gtrsim \text{several} \times 10^{22}$~cm$^{-2}$), consistently shows a high non-estimable ratio. This implies that a fraction of BH-LMXBs residing in or observed through these regions might be undetected by X-ray.

We note here that the detectability of a source does not directly correspond to our synthetic spectra being good enough (SNR $>$ 3) to model. This is because X-ray spectrum modelling requirements should be more stringent than a detection criterion. However, we believe that this finding should be highly correlated with the actual detectability.

\subsection{Bias-corrected observed galactic heights: implications on natal kicks and BH formation channels}

Now, we turn our attention back to the actual observed population. Figure~\ref{fig:height_distribution_vs_rho} confirms the findings of \citet{Jonker2021}, which indicate that the observed BH-LMXBs appear to be outside their expected formation region. The black holes in BH-LMXBs are expected to be correlated with regions where there is an abundance of massive progenitor stars (20 \solar~$\leq M_{\text{prog}} \leq$ 40 \solar) \citep{fryer2001}. These stars correlate highly with the thin-disc stellar population with a scale height of $\sim 20$\,pc \citep{reid2019}. It is argued in \citet{Jonker2021} that this discrepancy points to possible bias against detecting BH-LMXBs with higher masses due to observational selection effects at a lower $|z|$ than the scale height of massive stars. The bias arises from the weak natal kicks that massive BHs receive during their birth compared to the stronger ones that lower mass BHs receive \citep[e.g.][]{Repetto2012}. The stronger natal kicks would have the effect of propelling the BHs into higher $|z|$, effectively making them easier to detect compared to their massive counterparts, which remain in the galactic plane close to the location they were formed \citep{White1996,Jonker2021}.

We found that most sources have Galactic height $|z|$ less than 1\,kpc (Figure~\ref{fig:height_distribution}). This is higher than the theoretical expectations, but lower compared to the $|z|$ distribution of globular clusters, in particular, see Figure 2 in \citet[][]{Jonker2021}. Thus, our observed distribution provides yet another independent piece of evidence against the formation of BH-LMXBs in globular clusters.  

If we calculate the root-mean-squared (rms)-value of $z$ we find that when only considering \maxi~and \swift~sources, $z_{\text{rms}} =  969.53 \pm 447.09$\,pc, On the other hand, we find $z_{\text{rms}} = 864.43 \pm 267.41$\,pc, when we additionally include the \rxte~sources. The errors in these values are estimated using a bootstrap \citep{boostrap} resampling technique with 10,000 iterations. 
High rms Galactic height values have been argued to be evidence for high natal kick velocities during formation \citep[e.g.][]{Jonker2004,Atri2019,Jonker2021}. Since there is strong evidence that neutron star X-ray binaries, do receive high-velocity natal kicks \citep{Hobbs2005}, as evidenced by a $z_{\text{rms}}$ of $\sim 1000$\,pc \citep{Brandt1995,vanParadijs1995,White1996}. However, it is important to note that this would only be valid if there were no observational biases and large differences in formation evolution between neutron stars and BHs \citep{Repetto2017}.

Furthermore, an analysis by \citet{Atri2019} of the correlation between $|z|$ and their potential kick velocity distributions (PKV) suggests that a moderate correlation exists. However, they argue that concluding kick velocities through the $|z|$ distribution only, is not a reliable approach, because it does not explain some of the outliers the authors found in their data and the fact that these systems could be crossing the disc at the time of observation \citep{Atri2019}. In contrast, they state that $z_{\text{rms}}$ should provide a good indicator of the underlying population's natal kick magnitude, since it can, in principle, average out the time of observation bias \citep{Atri2019}.

Since we have found evidence of a hidden population at low galactic heights, we expect the currently observed rms value to be higher than the true population rms value. So, we believe that no firm conclusions should be drawn about high-velocity natal kicks solely from the current overall $z_{\text{rms}}$ estimate. 

\begin{figure}
    \centering
    \includegraphics[width=1\linewidth]{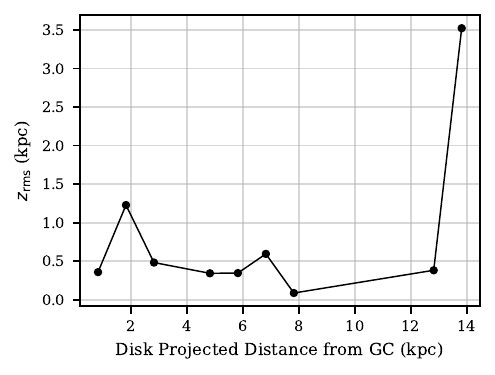}
    \caption{The observed disc radial variation of the root-mean-square (rms) height
        $z_{\text{rms}}$ from the galactic midplane. 
        The plot illustrates a roughly constant $z_{\text{rms}}$ within \(\sim10\) kpc, followed by a significant flare beyond 
        \(\sim12\) kpc. Compare to Figure 9 in \citep{Repetto2017}. We used the estimated median $z$ values and 1 kpc-wide bins.}
    \label{fig:z_rms_profile}
\end{figure}

In \citet{Repetto2017}, the authors investigate the relation between the rms height and the disc-projected radial distance using different binary population synthesis models that account for different formation scenarios. These formation scenarios correspond mainly to high or low natal kicks. They found that a model with some portion of BHs receiving high-velocity natal kicks is the best match to their observations sample \citep{Repetto2017}. However, regardless of the model, they found that the radial profile of $z_{\text{rms}}$ had an increasing trend with increased radial distance. We leveraged our estimations to probe the disc-projected radial profile of $z_{\text{rms}}$. We plot such a graph in Figure~\ref{fig:z_rms_profile} with 1 kpc-wide radial bins and using the full sample with \maxi, \swift, and \rxte. Our rms values are mostly around 0.5 kpc, except for the abrupt increase to $\sim 3.5$\,kpc at a radius of $\sim 13$\,kpc (this is only due to one source). So, our profile does not suggest a gradual $z_{\text{rms}}$ increase as what was found using the binary population synthesis models in \citep{Repetto2017}. We caution here, that although we have a relatively large number of BH-LMXBs compared to previous population studies, we still may not have a sufficient number of them to produce a true representation of the radial profile of $z_{\text{rms}}$ given the 1\,kpc bins, as well as the evidence for a hidden population. Furthermore, we also note that we did not consider the uncertainty of $z$ or Galactic radius estimations arising from the distance uncertainty. A more robust method would be to perform Monte Carlo sampling from the distance distribution for each source and quantify both uncertainties in each bin. We do argue, however, that this is not likely to cause major changes in the profile given the relatively low uncertainties of the radius, and $z$ (see Figure~\ref{fig:height_distribution_vs_rho}).

There has been plenty of evidence provided in favour of both high natal kicks \citep[e.g.][]{Fargos2009,Repetto2012,Repetto2017,Atri2019,Kimball2023,MataSanchez2025}, and low natal kicks \citep[e.g.][]{Mandel2016,Wang2024,Janka2024,Burdge2024,Nagarajan2024} for BH-LMXBs. Most recently, the discovery that V404 Cygni is, in fact, a triple system \citep{Burdge2024}, provides one of the most concrete evidence for low natal kicks. As part of their analysis, the authors analysed the velocities of nearby stars and concluded that the peculiar velocity of V404 Cygni is not significantly different \citep{Burdge2024}. Given this and the tertiary discovery, they estimate the natal kick velocity as $\lesssim 5$\,km\,s$^{-1}$ \citep{Burdge2024}. Furthermore, a simulation study \citep{Shariat2024}, inspired by the V404 Cygni discovery, was conducted by utilising the state-of-art codes that incorporate triple dynamics, stellar evolution, and natal kicks. They additionally compare their triple population synthesis models with isolated binary synthesis models \citep{Shariat2024}. Their simulations strongly suggested that at least some BH-LMXBs form with very low natal kicks and that the triple formation channel is more efficient in producing BH-LMXBs than the isolated binary channel \citep{Shariat2024}. Besides this simulation study, the strong evidence of low natal kick in V404 Cygni, along with a new analysis of BH sources using the 3rd data release from \textit{Gaia} \citep{Gaia}, prompted \citet{Nagarajan2024} in their recent investigation to favour a hybrid origin model where some BHs gain high kicks from supernova explosions, while other BHs receive minimal kicks. 

As a final note, our observed $z_{\text{rms}}$ and its uncertainties, combined with our evidence towards a hidden population in low galactic heights, seem to be most compatible with a wide range in the natal kick distribution and hence with mixed origins of BH-LMXBs, where some systems receive high kicks and some receive no or very low kicks, consistent with the most recent studies \citep{Shariat2024,Nagarajan2024}.

\section{Conclusions}\label{sec:conclusions}

In this work, we revisited the spatial distribution of Galactic black hole low-mass X-ray binaries (BH-LMXBs) by leveraging a new and fully independent set of distance measurements obtained from soft-state and, if applicable, soft-to-hard transition X-ray spectral state modelling, developed in our previous work \citep{Abdulghani2024}. One of our goals in this current study was to mitigate systematic instrument biases stemming from variations in interstellar absorption, BH mass/spin, disc temperature, exposure time, and inclination by utilising simulations. Another goal was to gain some insight into observational selection effects through the simulation results. Finally, the main aim was to present an independent bias-corrected view of the spatial distribution of a large sample of Galactic BH-LMXBs. Our findings can be summarised as follows:

\begin{enumerate}
    \item \textbf{Soft-state distance method and bias correction.} 
    Our large grid of spectral simulations demonstrates that distance estimates derived from thermal disc-dominated states (i.e.\ the ``soft state'') can suffer from systematic bias, especially for sources at large heliocentric distances. We have shown that the method is generally robust when the disc-to-total flux ratio is high ($\sim 0.9$), the maximum disc temperature is $\geq$ 1\,keV, and the black hole is not rotating. However, we have also found that the method can be unreliable when the disc-to-total flux ratio is low $\sim 0.2$, the maximum disc temperature $\leq$ 0.5\,keV, or the black hole is maximally spinning. We quantified a general bias trend and used it as a bias corrector.

    \item \textbf{Evidence for a hidden population.}
    After applying bias corrections, the disc-projected radial distribution of BH-LMXBs is broadly consistent with the high-mass stellar distribution models of the Milky Way \citep{Grimm2002}. Nonetheless, our simulations and observed distributions highlight the likelihood of missing systems close to the Galactic plane and in crowded fields like near the Galactic Centre, where interstellar absorption is highest. This implies that while a central population might possess a lower scale height, its highly absorbed BH-LMXBs, at low $|z|$, may remain undetected, driving a selection effect that artificially inflates the observed mean-scale height above the plane.

    \item \textbf{Galactic height and natal kicks.}
    A key open question in black hole formation is whether black hole natal kicks are generally large or small \citep[see e.g.][]{Repetto2017,Atri2019,Nagarajan2024}. We find that the root-mean-squared (rms) height of BH-LMXBs in our sample lie within $z_{\text{zms}} \approx 0.5\text{--}1.3\,\mathrm{kpc}$ of the Galactic plane. This is higher than the small-scale height of high-mass star-forming regions in which black holes presumably originate ($\sim 0.02$ kpc), but still lower than the heights typically associated with globular clusters \citep[e.g.][]{Jonker2021}. While this observed $z_{\text{rms}}$ could be influenced by the aforementioned selection effects against low-$|z|$ and inner Galaxy sources, therefore, we cannot definitively confirm whether BH natal kicks are large, small, or both (bi-modal). We believe that our results hint that scenarios that consider a mixed origin of black hole formation with small and high kicks would be the most consistent with observations. Nevertheless, a more complete sample at small $|z|$ is needed for robust natal-kick constraints.

    \item \textbf{Future work.}
    Upcoming and ongoing all-sky monitoring missions, as well as deeper follow-up in soft X-rays, will help alleviate absorption biases. Improved sensitivity at low energies (such as \textit{ATHENA} and \textit{AXIS}) will be critical, as it enhances the disc-dominated flux recovery and, thus, the accuracy of distance determination from the soft-state spectral method. Detection and distance determination through sensitive hard X-ray telescopes (e.g. \textit{STROBE-X}) could also play an important role in alleviating the challenge.  While these advancements in X-ray capabilities are expected to significantly mitigate the current observational biases against the highly absorbed, low-$|z|$ central population, thereby providing a clearer basis for differentiating natal kick scenarios, they will be powerfully complemented by other approaches. Specifically, upcoming facilities like the Nancy Grace Roman Space Telescope, through its microlensing survey of the Galactic Bulge \citep{Sajadian2023}, and the Square Kilometre Array \citep{Schoedel2024}, with its ability to penetrate dust and provide high-resolution radio astrometry, will offer crucial, independent pathways to detect and characterize BH and BH-LMXBs in these obscured regions. The extreme source crowding and peak interstellar absorption towards the innermost Galactic regions suggest that fully overcoming this challenge with any single method to reveal the entirety of the central BH-LMXB population may remain a persistent difficulty. However, the synergy of these multi-wavelength and multi-technique efforts (including X-ray, radio, gravitational waves and microlensing) will provide a far more complete picture than is currently possible.
\end{enumerate}

We emphasise that our analysis here has not included all $\sim 70$ \citep{Corral-Santana2016} currently detected BH-LMXBs. Notably, systems exhibiting failed outbursts (hard-only states) were excluded. This omission introduces a potential bias towards systems more likely to transition clearly between spectral states, possibly affecting our derived spatial distributions and the inferred natal-kick conclusions. To overcome this limitation, future studies should incorporate the full diversity of outburst types by developing robust distance estimation techniques applicable even in the absence of distinct state transitions. Although the majority of BH-LMXBs do exhibit the typical behaviour that the systems included in our investigation do, including such systems could slightly modify our understanding of the Galactic distribution and potentially reveal a more comprehensive picture of BH formation mechanisms.

In general, our findings underscore the significance of selection effects when constraining BH-LMXB demographics. While X-ray spectral fits provide a valuable method for distance estimation, the resulting distributions must be treated with care. Future dedicated surveys and follow-up programs, combined with simulations that account for realistic absorption and source evolution, will be crucial for progressing toward a comprehensive picture of black hole formation in the Milky Way.


\section*{Acknowledgements}

We appreciate the careful review and the highly helpful suggestions provided by the anonymous reviewer that enhanced the quality of this paper. JC acknowledges support from the Leverhulme Trust grant RPG-2023-240. Computational efforts were performed on the Tempest High-Performance Computing System, operated and supported by University Information Technology Research Cyberinfrastructure (RRID:SCR\_026229) at Montana State University. This research leveraged the Astropy package:\footnote{\url{http://www.astropy.org}} a community-developed core Python package and an ecosystem of tools and resources for astronomy \citep{astropy:2013, astropy:2018, astropy:2022}. 

\section*{Data Availability}
The scripts used to generate the simulations and the simulations results (as an SQLite database) can be found in the following repository: \url{https://github.com/ysabdulghani/xrb-population}. Tables of the corrected distance estimations can also be found at the same link.



\bibliographystyle{mnras}
\bibliography{reference} 




\appendix

\section{Error distributions}\label{app:distributions}

The simulation done in section~\ref{sec:simulation-study}, produced thousands of results as mentioned in that section. If you look at only one parameter value from one of the eight, there are still $\sim 1000$ other combinations. In this section, we explore how does the error distributions, when looking at individual parameter values, change with input distance. 

As pointed out in the main text, the following transformation was applied to the deviations $\Delta D_{\text{est,in}}$:
\begin{equation}
 \text{log-mod($\Delta D_{\text{est,in}}$)}=\text{sign($\Delta D_{\text{est,in}}$)} \log (|D_{\text{est,in}}|+1)
 \label{eqn:log-modulus}
\end{equation}

\begin{figure}
    \includegraphics[width=0.95\linewidth]{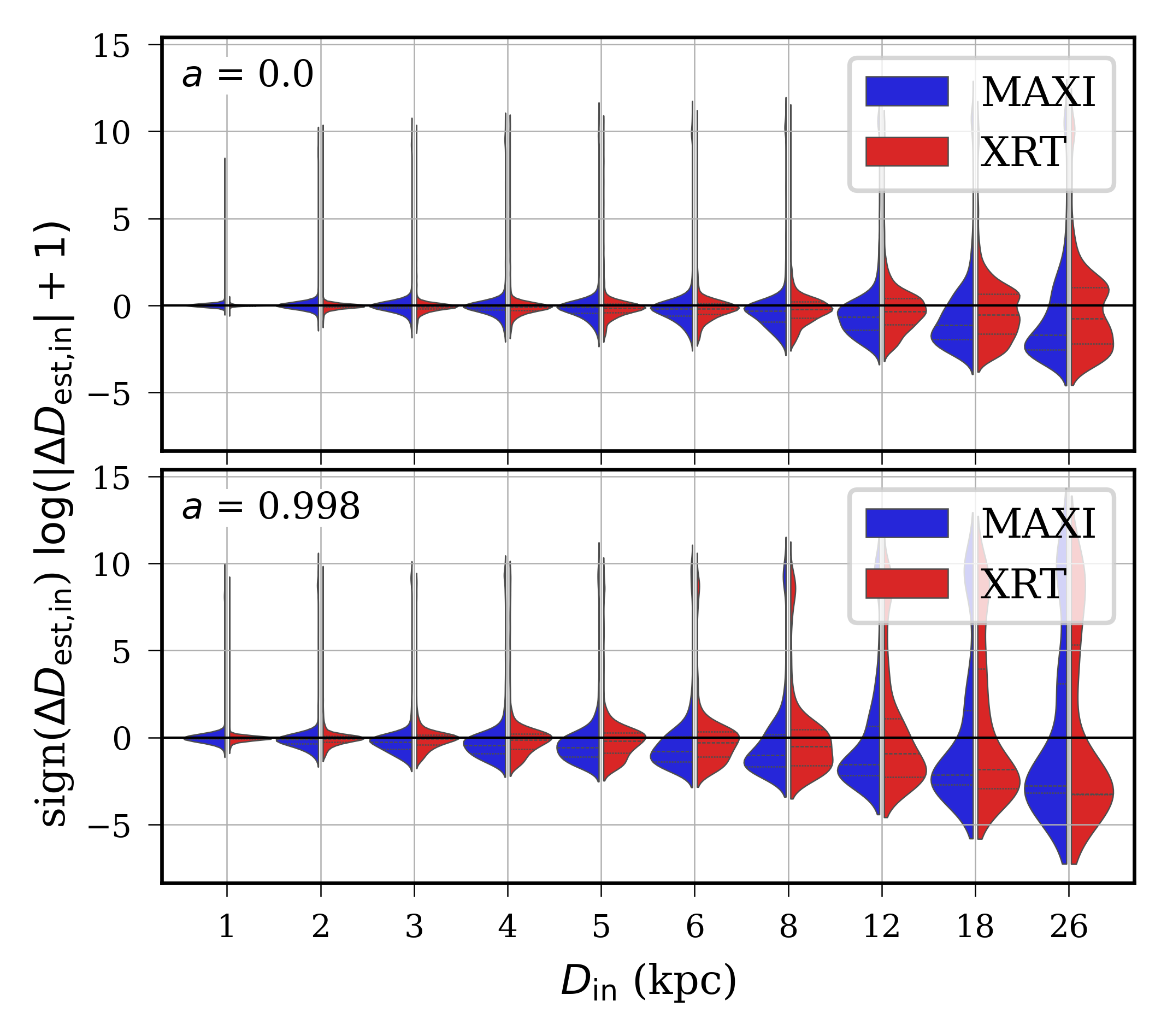}
    \caption{This shows the distributions of the log-modulus transformation of the error $\Delta D_{\text{est,in}}$ for the different BH spin $a$ simulation values after fitting the synthetic spectra. The horizontal solid lines show where $\Delta D_{\text{est,in}} = 0$.}
    \label{fig:a-dist}
\end{figure}

It is clear from Figure~\ref{fig:a-dist} that the black hole's spin significantly influences the ability of the method to retrieve the input distance value accurately. The error distribution is wider for the high-spin ($a=0.998$) case. Furthermore, the peaks of the error distributions of the high-spin case for $D_{\text{in}} \geq 6$ kpc lie at considerably lower values than the zero-error line. Both \maxi/GSC and \swift/XRT results show a heavy right tail in the distribution for the high-spin case, especially at high distances. Additionally, the zero-spin simulations for \swift/XRT show bimodal distributions of the error compared to the high-spin case at a high distance $> 12$ kpc. The peaks of the \swift/XRT bimodal distributions are less pronounced but are closer to the zero line than the \maxi/GSC distributions. The zero-spin case for \swift/XRT simulations appears to yield peaks that are very close to the input distance, even at the highest input distances. The \maxi/GSC results, on the other hand, show peaks that shift lower as the input distance increases.

\begin{figure}
    \includegraphics[width=0.95\linewidth]{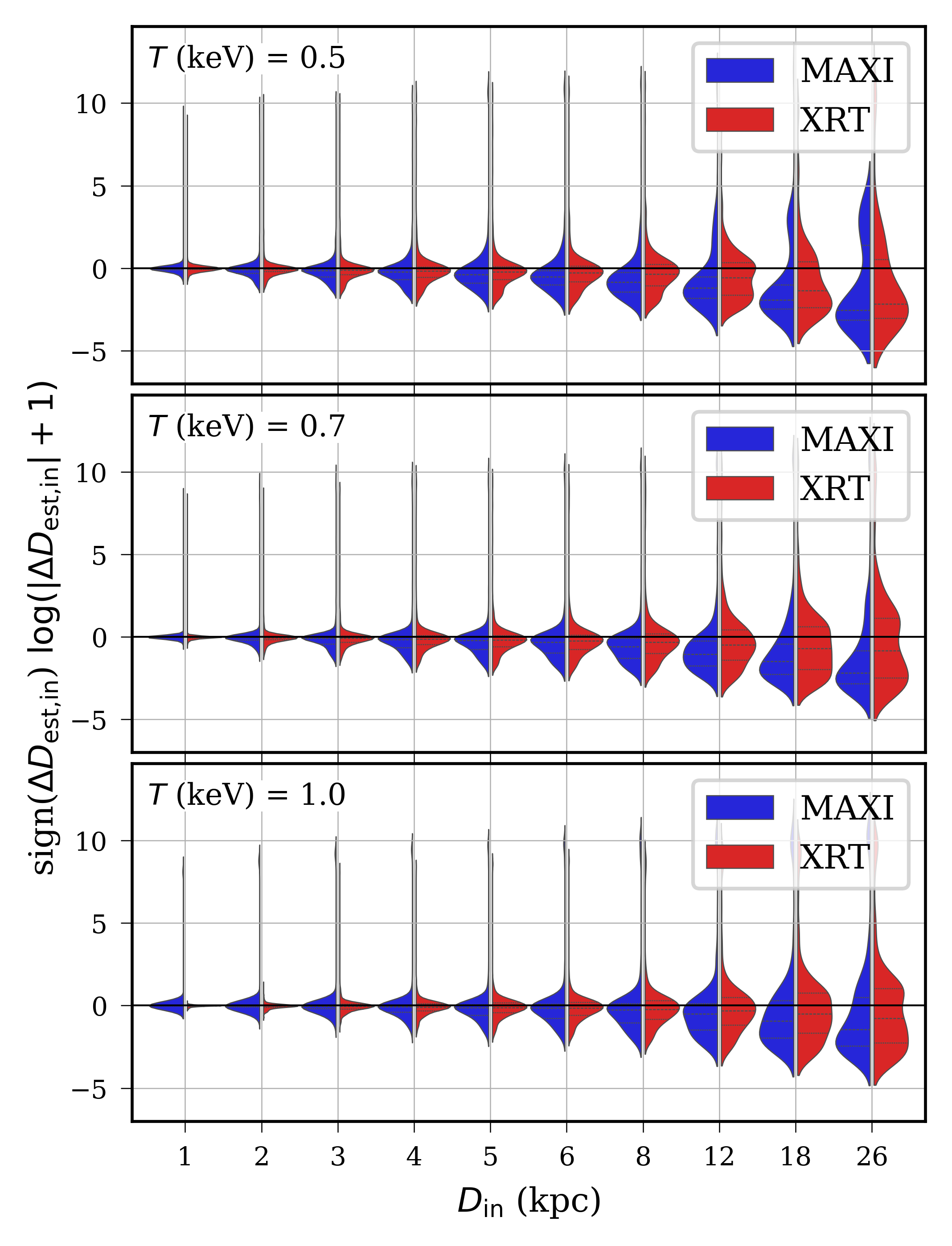}
    \caption{This shows the distributions of the log-modulus transformation of the error $\Delta D_{\text{est,in}}$ for the different maximum disc temperature $T$ simulation values after fitting the synthetic spectra. The horizontal solid lines show where $\Delta D_{\text{est,in}} = 0$.}
    \label{fig:T-dist}
\end{figure}

It can also be observed from Figure~\ref{fig:T-dist} that the maximum disc temperature $T$ is another highly influential model parameter. Overall, if the maximum disc temperature is lower, we obtain higher error values. Nevertheless, there is an exception to that trend at the 1\,kpc and 2\,kpc distributions, which appear to be narrower at $T=0.7$ keV for \maxi/GSC only.

\begin{figure}
    \includegraphics[width=0.95\linewidth]{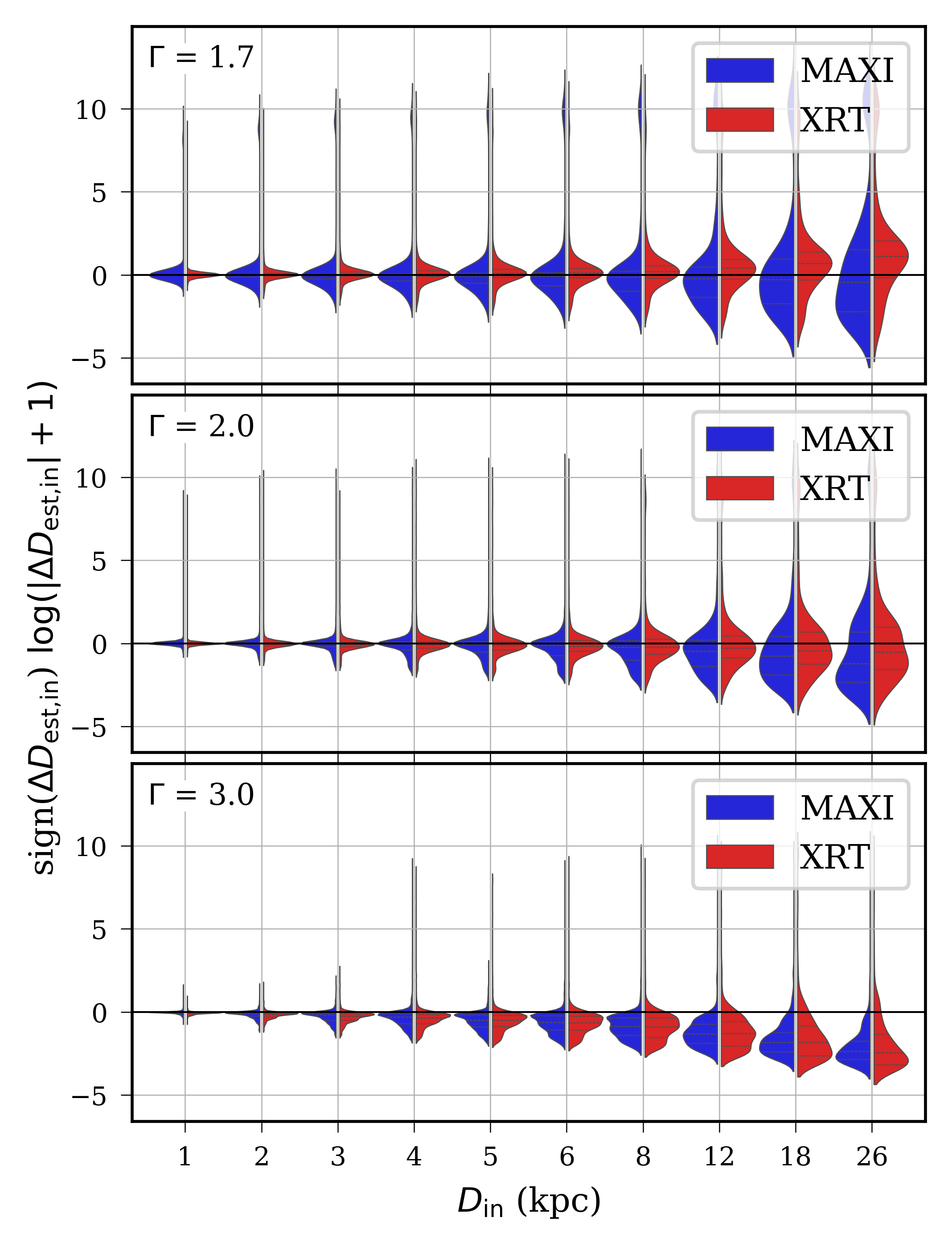}
    \caption{This shows the distributions of the log-modulus transformation of the error $\Delta D_{\text{est,in}}$ for the different power law index values $\Gamma$ simulation values after fitting the synthetic spectra. The horizontal solid line shows where $\Delta D_{\text{est,in}} = 0$ is.}
    \label{fig:g-dist}
\end{figure}

The distributions of the errors for the different $\Gamma$ values, as shown in Figure~\ref{fig:g-dist}, clearly reveal another high influence parameter. In particular, two major effects are evident. The distribution is wider at lower $\Gamma$ and narrows as it increases. The other effect is that the entire distribution, along with its peaks, shifts lower as $\Gamma$ increases. This corresponds to an increase in the magnitude of the distance underestimation. Another interesting observation is the disappearance of the very high error outliers at the highest $\Gamma$ value, where input distances are $\leq 4$\,kpc for both instruments.

\begin{figure}
    \includegraphics[width=0.95\linewidth]{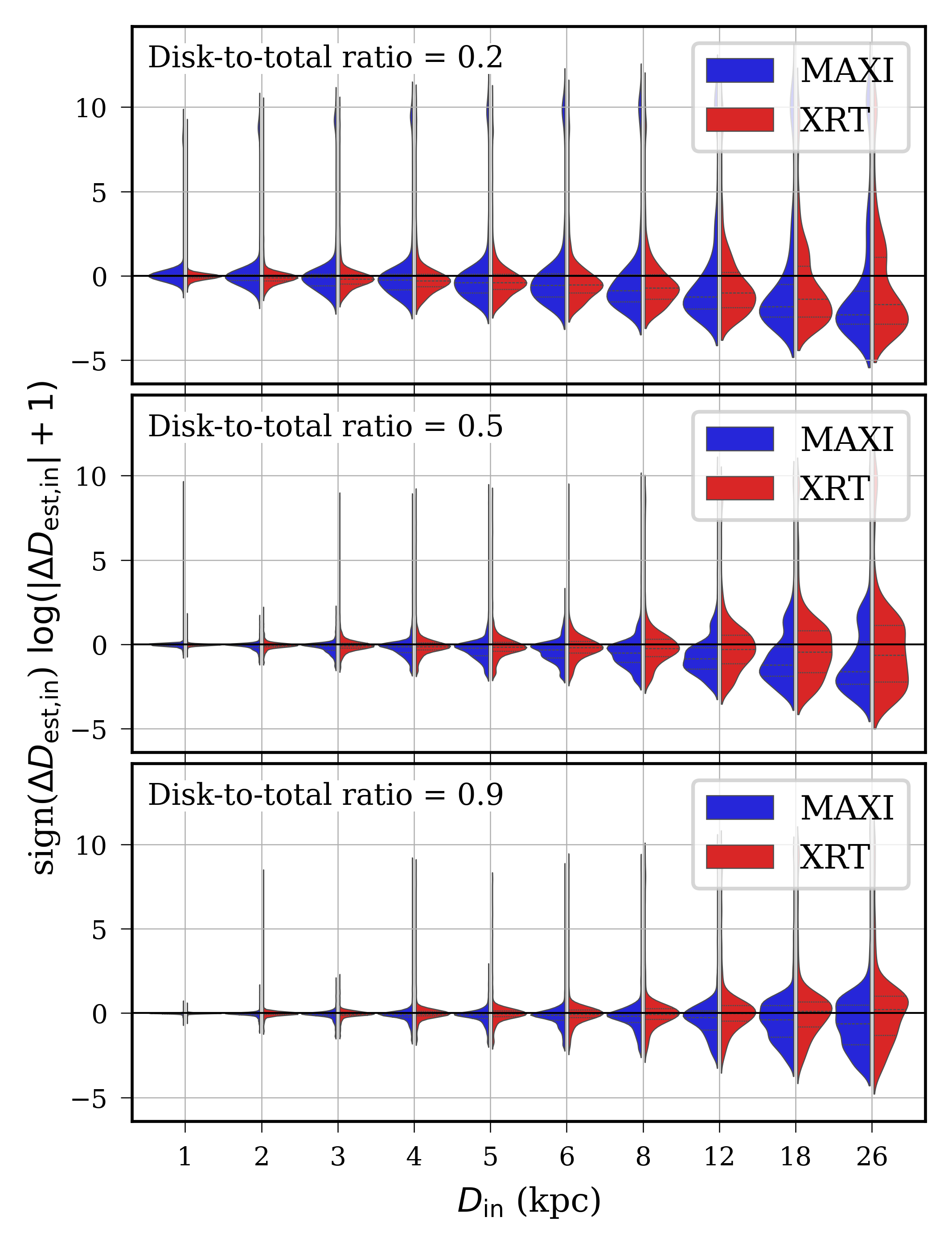}
    \caption{This shows the distributions of the log-modulus transformation of the error $\Delta D_{\text{est,in}}$ for the different disc-to-total ratio simulation values after fitting the synthetic spectra. The horizontal solid lines show where $\Delta D_{\text{est,in}} = 0$.}
    \label{fig:r-dist}
\end{figure}

The last parameter that has a high impact on the errors is the disc-to-total ratio. As apparent from Figure~\ref{fig:r-dist}, the peak of the distributions is much closer to the zero line for the highest ratio. This is, of course, expected given the method dependence on modelling the disc flux. The width of the distributions, however, shows an interesting pattern for \maxi/GSC, where it generally shrinks as the ratio increases. But for \swift/XRT, it is the widest at the middle ratio value. 

\begin{figure}
    \includegraphics[width=0.95\linewidth]{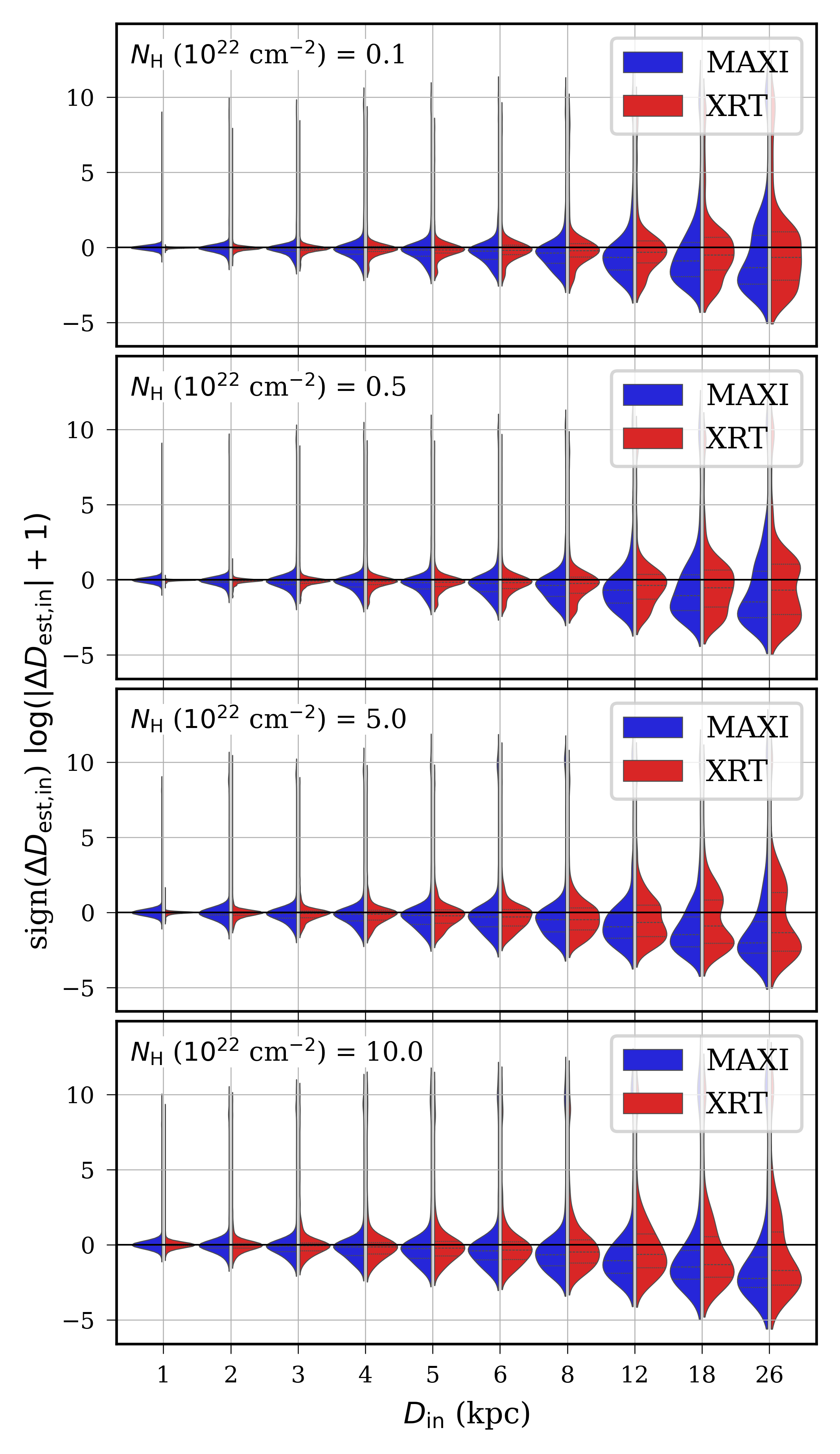}
    \caption{This shows the distributions of the log-modulus transformation of the error $\Delta D_{\text{est,in}}$ for the different $N_{\text{H}}$ simulation values after fitting the synthetic spectra. The horizontal solid lines show where $\Delta D_{\text{est,in}} = 0$.}
    \label{fig:nH-dist}
\end{figure}

Figure~\ref{fig:nH-dist} shows the distribution of the log-modulus errors that resulted from the modelling of the simulated spectra for each of the grid points of the 10$\times$4 $D_{\text{in}}$-$N_{\text{H}}$ grid for \maxi/GSC and \swift/XRT. As expected, the resulting error distribution is narrow when the distance is close (1-3\,kpc) but widens considerably as the input distance and $N_{\text{H}}$ increases. Overall, the distributions appear to be left-skewed, indicating that the error in the measured distances is more likely to be an underestimation. There are also several outliers with very high and very low estimated distances, as evidenced by the plot scale. Comparing the \maxi/GSC to \swift/XRT results, it is clear that the lower distance distributions (1-3\,kpc) are generally narrower for \swift/XRT results, except in the $N_{\text{H}} = 10^{23}$\,cm$^{-2}$ case. Moreover, the distributions at higher input distances $>8$ kpc show wider distributions with an apparent extra modality in the \swift~results compared to \maxi~results. However, their median values appear to be closer to the zero-error line.

\begin{figure}
    \includegraphics[width=0.95\linewidth]{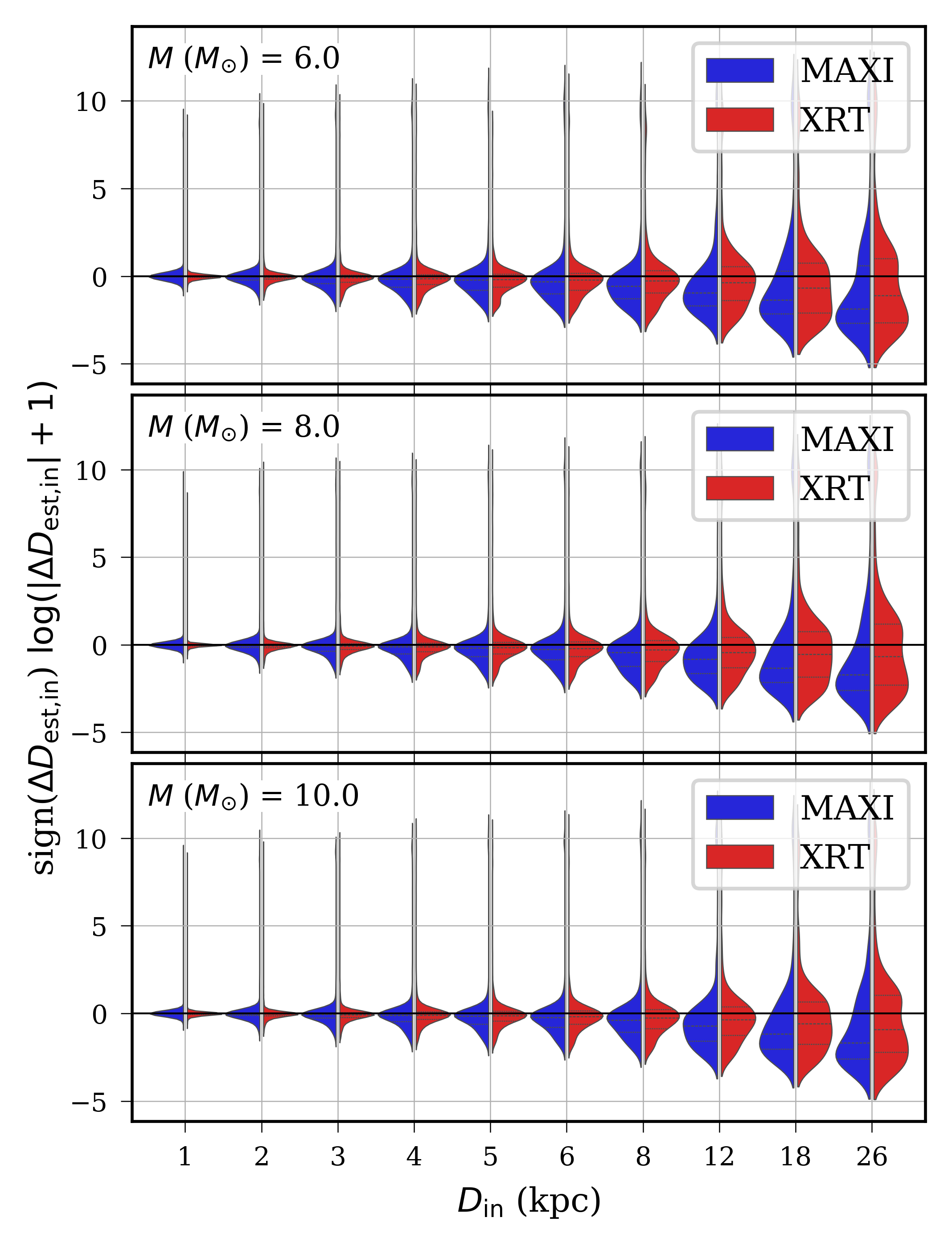}
    \caption{This shows the distributions of the log-modulus transformation of the error $\Delta D_{\text{est,in}}$ for the different BH mass $M$ simulation values after fitting the synthetic spectra. The horizontal solid lines show where $\Delta D_{\text{est,in}} = 0$.}
    \label{fig:m-dist}
\end{figure}

\begin{figure}
    \includegraphics[width=0.95\linewidth]{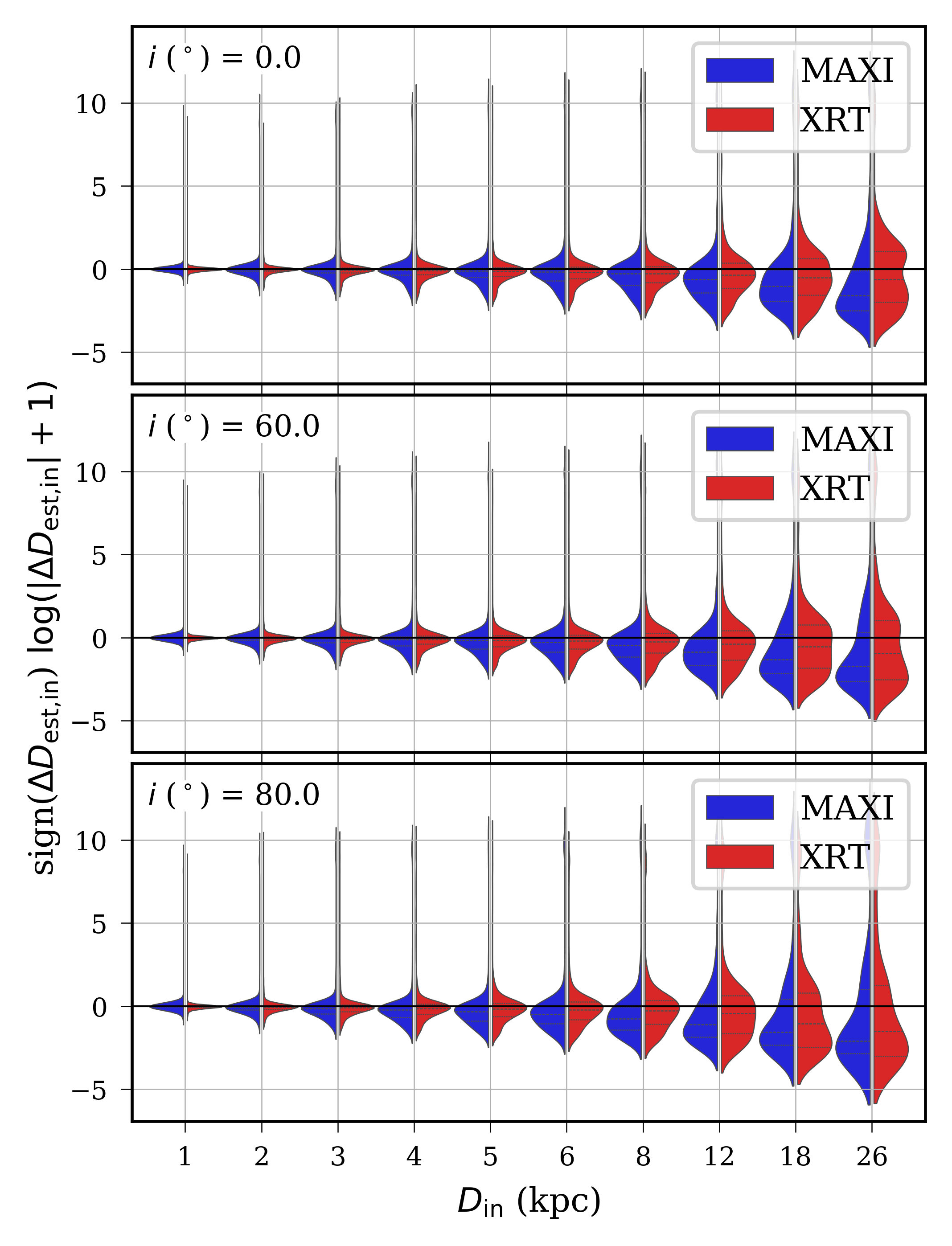}
    \caption{This shows the distributions of the log-modulus transformation of the error $\Delta D_{\text{est,in}}$ for the different disc inclination $i$ simulation values after fitting the synthetic spectra. The horizontal solid lines show where $\Delta D_{\text{est,in}} = 0$.}
    \label{fig:i-dist}
\end{figure}

\begin{figure}
    \includegraphics[width=0.95\linewidth]{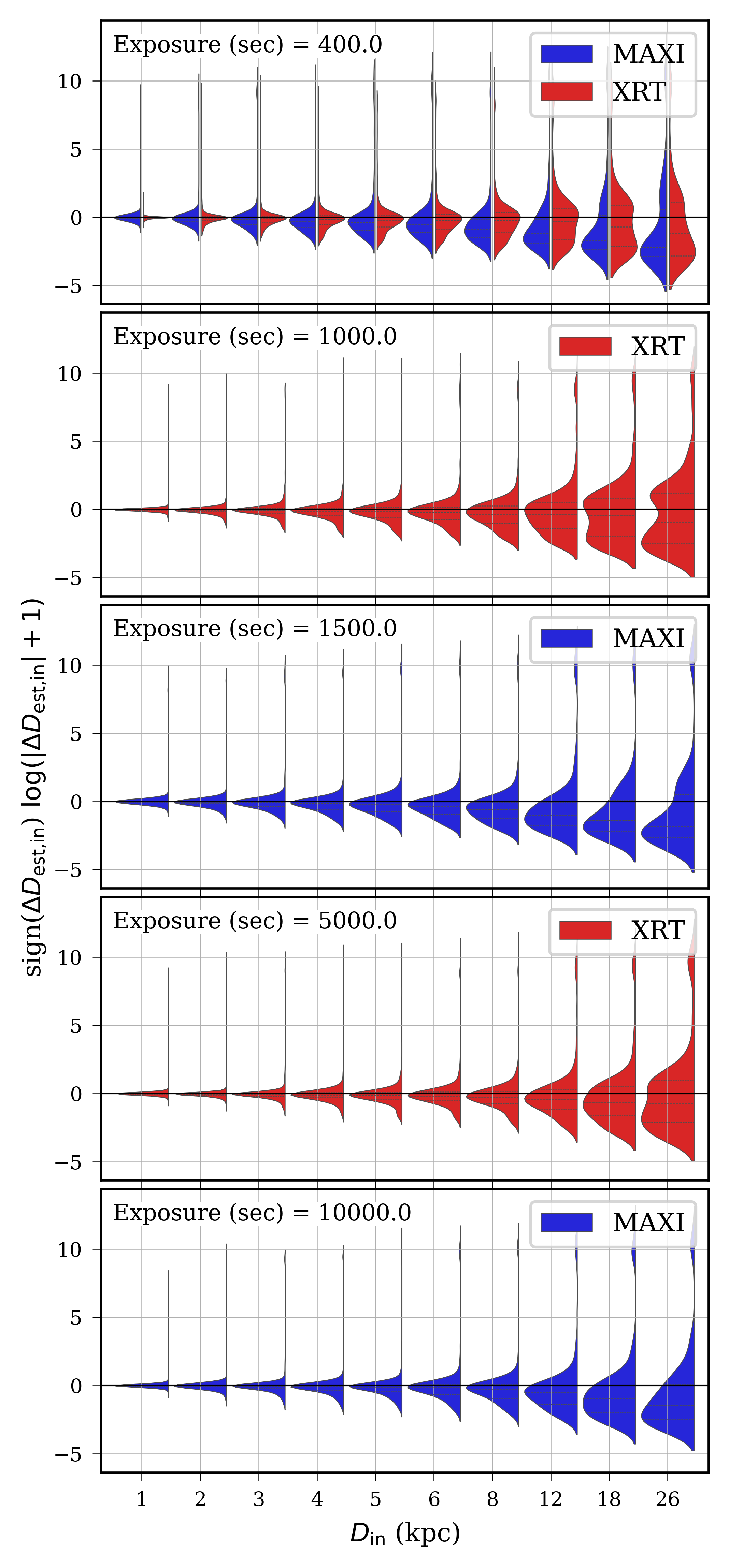}
    \caption{This shows the distributions of the log-modulus transformation of the error $\Delta D_{\text{est,in}}$ for the different exposure time simulation values after fitting the synthetic spectra. The horizontal solid lines show where $\Delta D_{\text{est,in}} = 0$.}
    \label{fig:e-dist}
\end{figure}

Finally, we show the distributions for the BH mass $M$, disc inclination, and the exposure time in Figures~\ref{fig:m-dist}, \ref{fig:i-dist}, and~\ref{fig:e-dist}, respectively. These parameters show moderate to low effects on the distribution shape, with the changes in $M$ having the weakest effect on the error. 

\section{Additional interaction effects figures}\label{app:interactions}

\begin{figure}
    \includegraphics[width=0.95\linewidth]{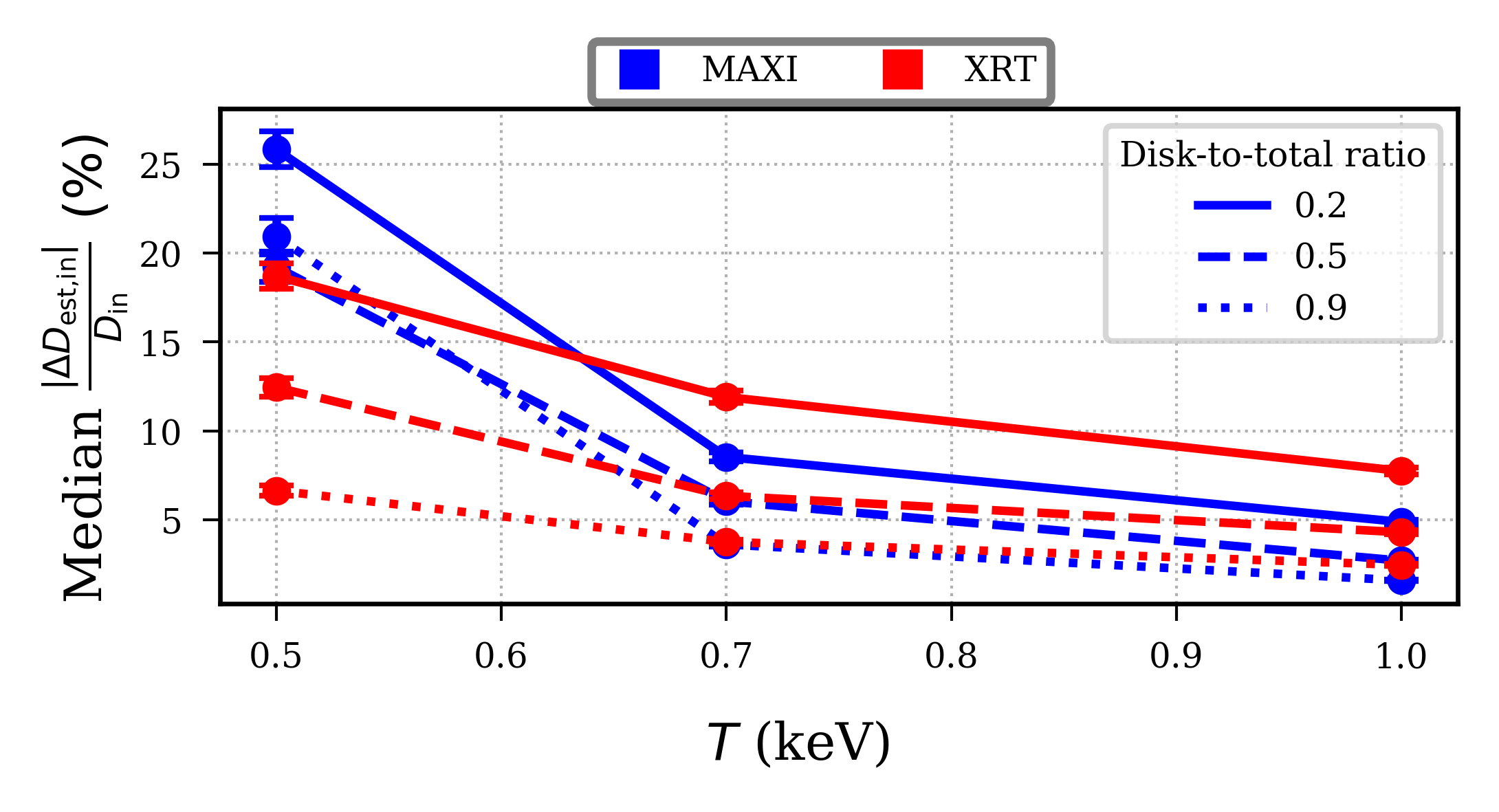}
    \caption{This figure shows pairwise interactions between $T$ and the disc-to-total ratio. These interactions were selected via the backward stepwise BIC process.}
    \label{fig:T-interactions1}
\end{figure}

\begin{figure}
    \includegraphics[width=0.95\linewidth]{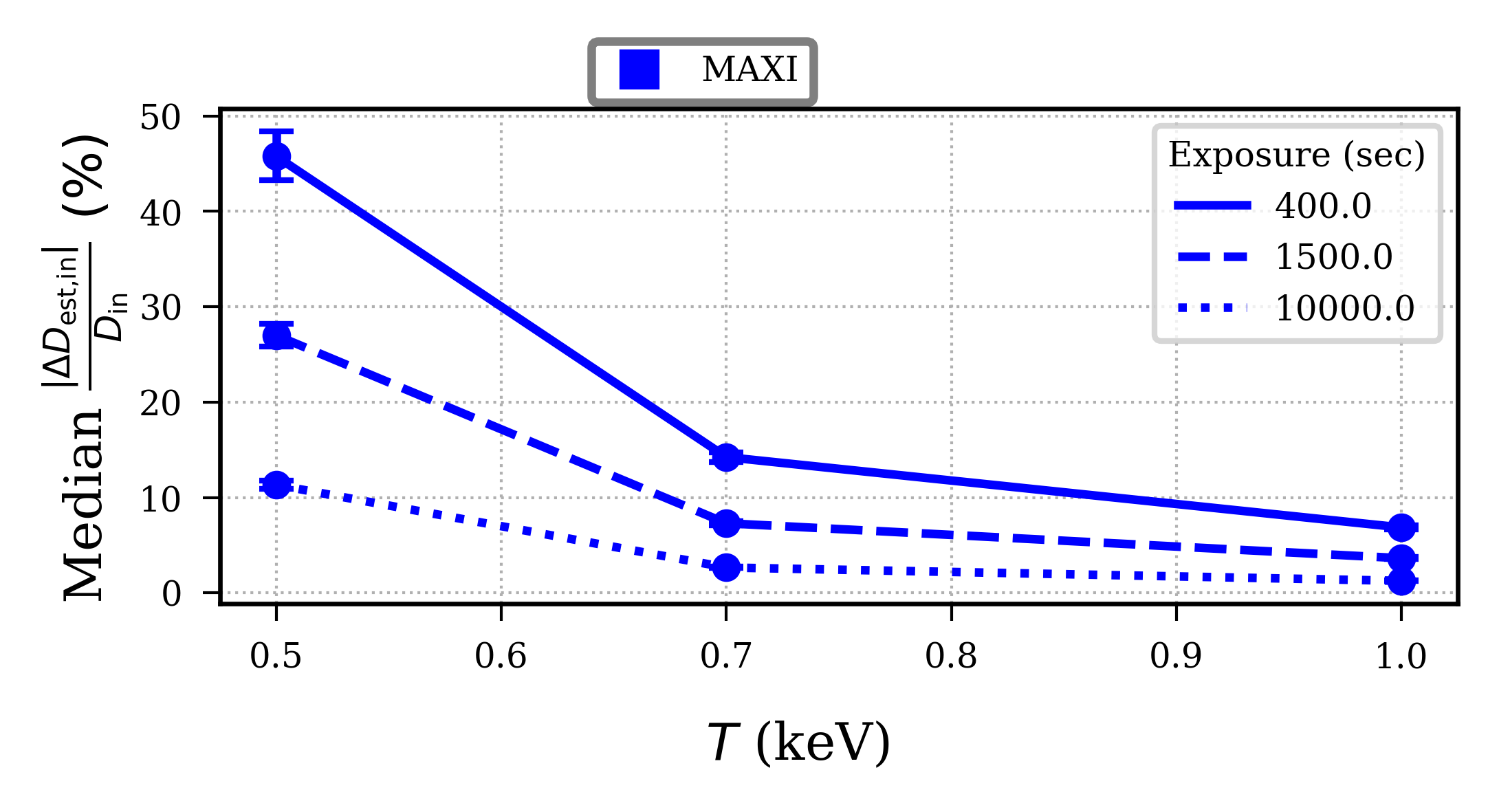}
    \caption{This figure shows pairwise interactions between $T$ and the exposure time. These interactions were selected via the backward stepwise BIC process. This interaction was only selected for \maxi/GSC results.}
    \label{fig:T-interactions2}
\end{figure}

\begin{figure}
    \includegraphics[width=0.95\linewidth]{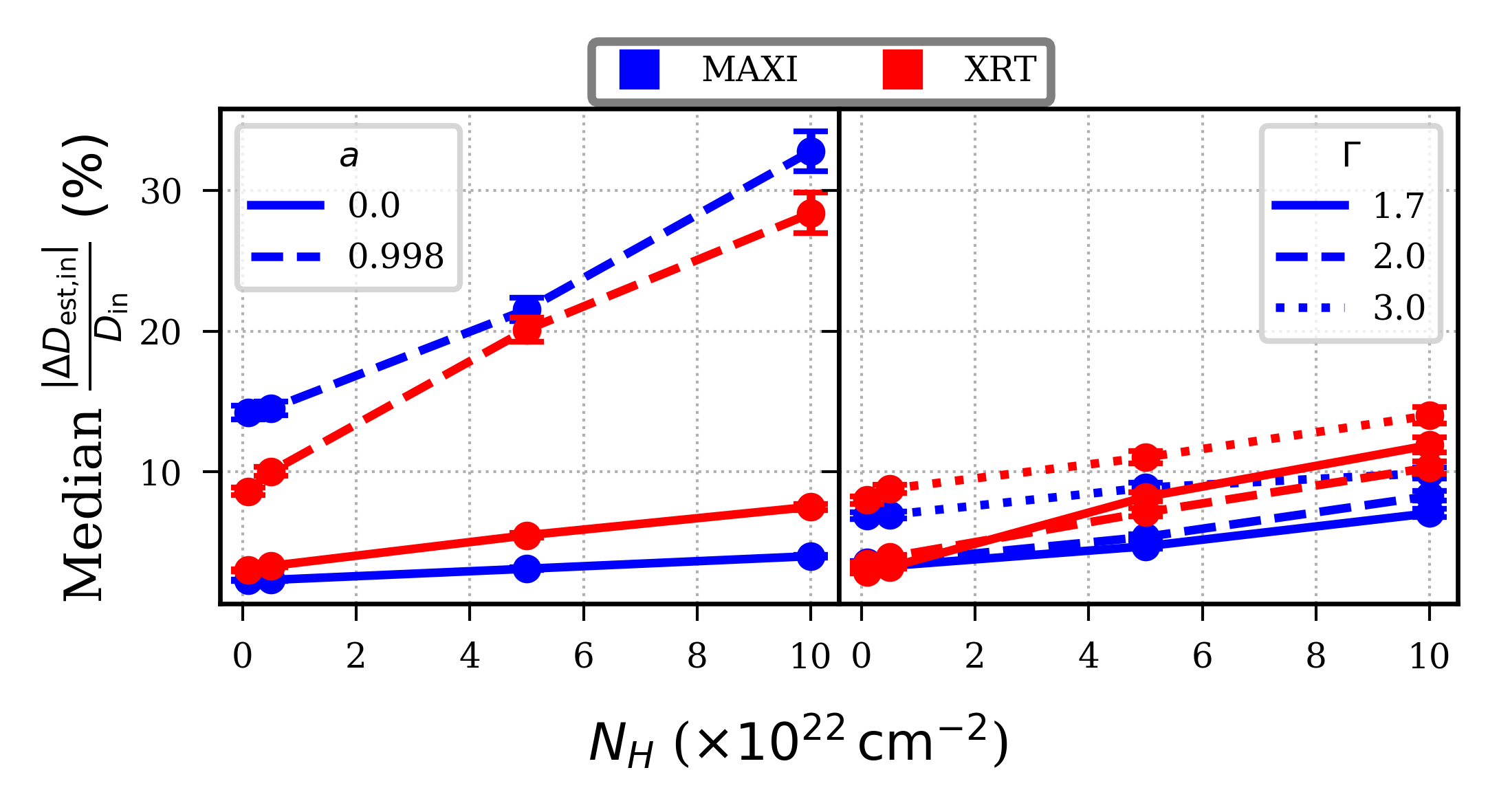}
    \caption{This figure shows pairwise interactions between $N_{\text{H}}$ and other parameters. These interactions were selected via the backward stepwise BIC process.}
    \label{fig:nH-interactions}
\end{figure}

\begin{figure}
    \includegraphics[width=0.95\linewidth]{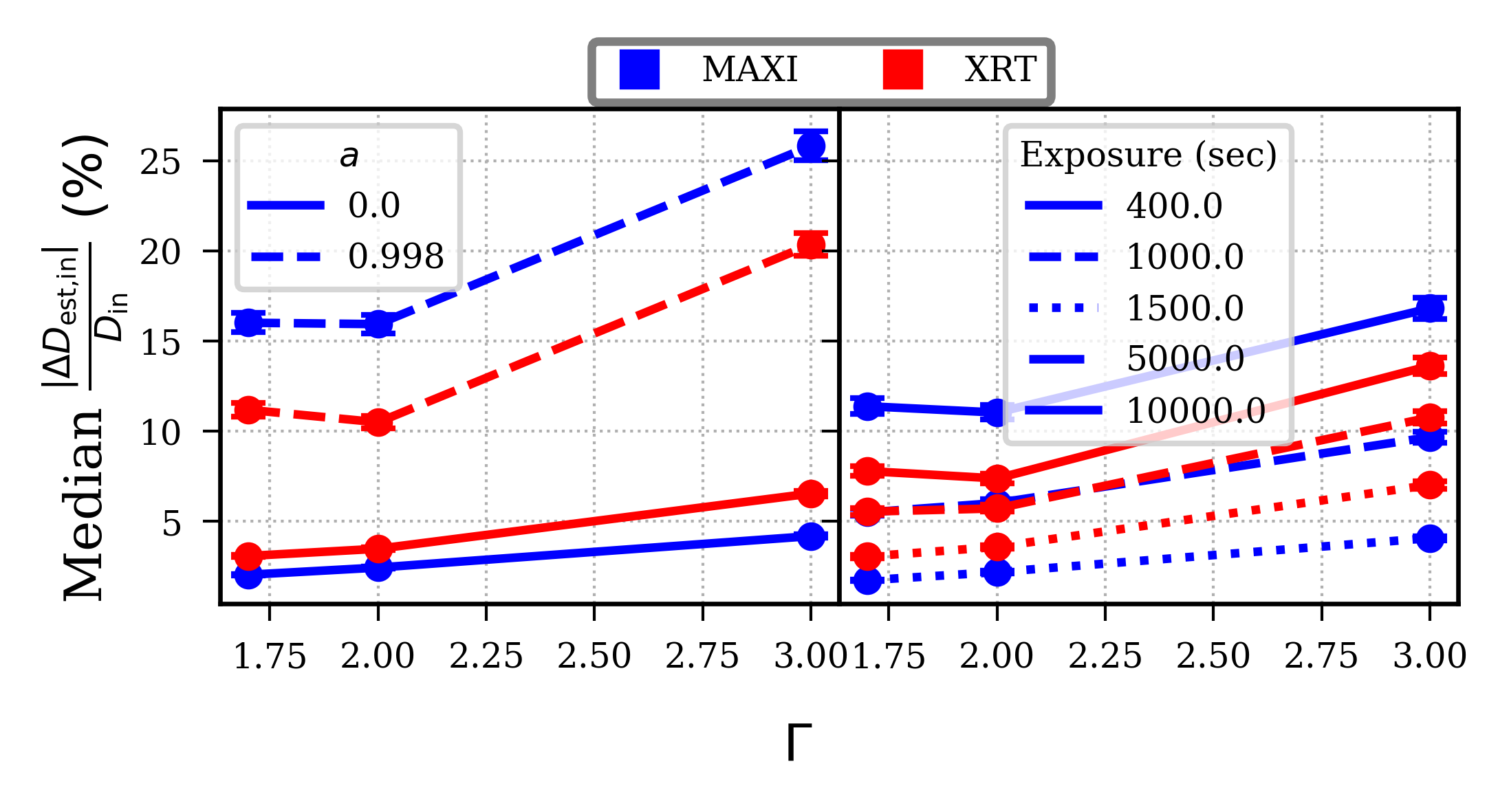}
    \caption{This figure shows pairwise interactions between the $\Gamma$ and other parameters. These interactions were selected via the backward stepwise BIC process.}
    \label{fig:g-interactions}
\end{figure}

\begin{figure}
    \includegraphics[width=0.95\linewidth]{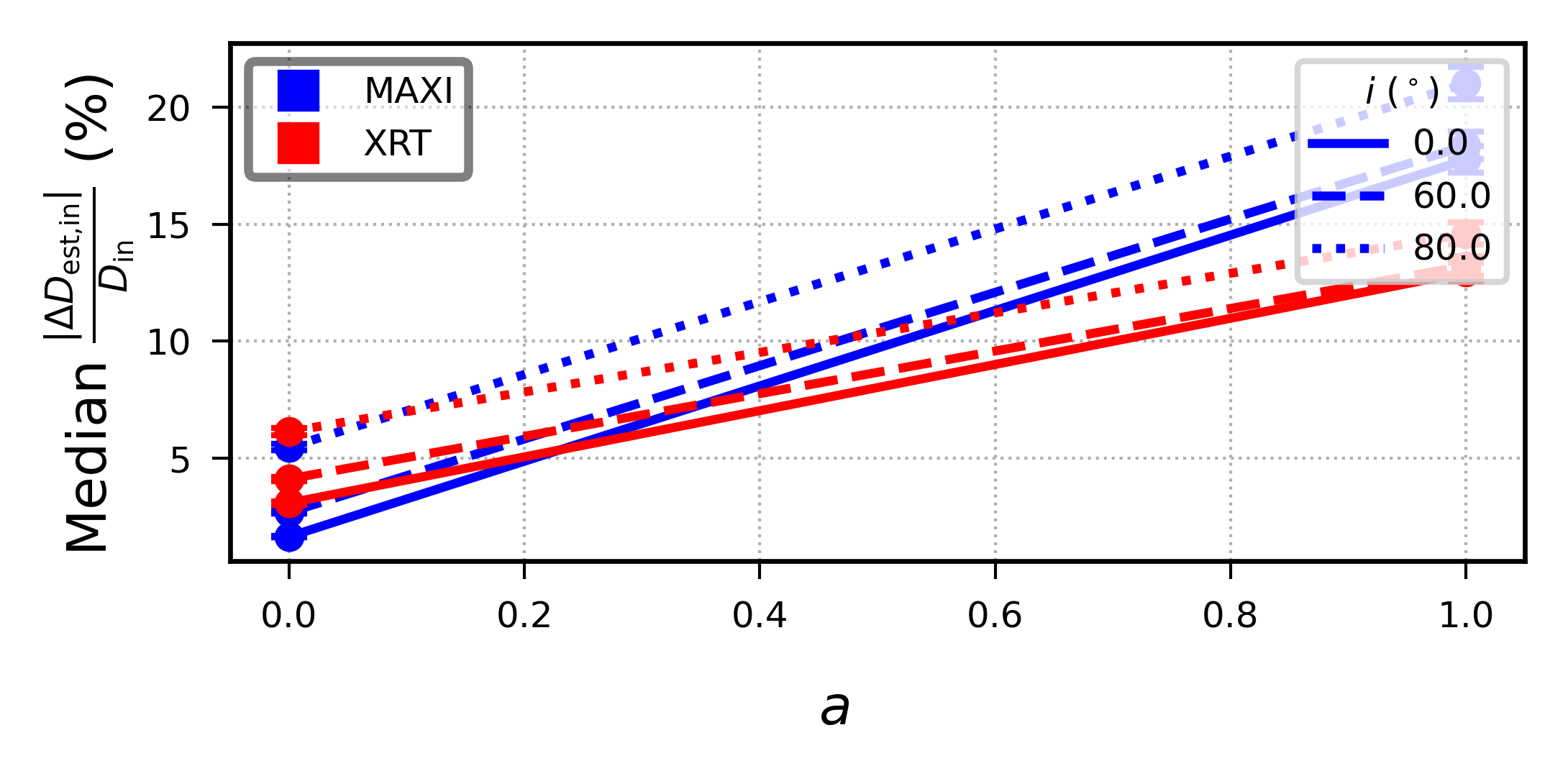}
    \caption{This figure shows pairwise interactions between $a$ and the inclination. These interactions were selected via the backward stepwise BIC process.}
    \label{fig:a-interactions1}
\end{figure}

\begin{figure}
    \includegraphics[width=0.95\linewidth]{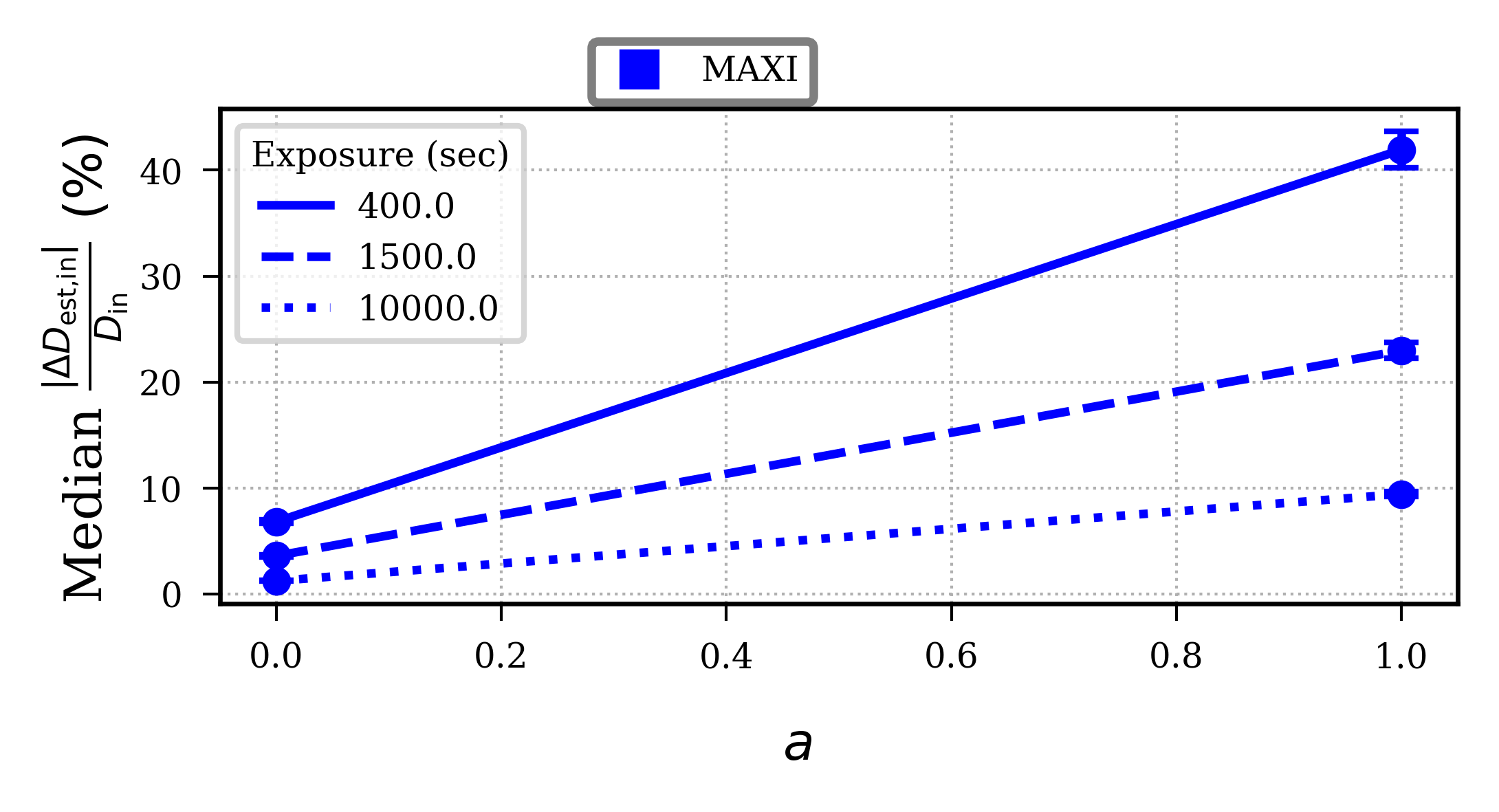}
    \caption{This figure shows pairwise interactions between $a$ and the exposure time. These interactions were selected via the backward stepwise BIC process. This interaction was only present in the final\maxi/GSC model.}
    \label{fig:a-interactions2}
\end{figure}
 
\section{Candidate functions for fitting the accurate distance distribution}

\begin{table}
\centering
\caption{Forms of the candidate probability density functions (PDFs)}
\label{tab:pdfs}
\begin{tabular}{@{}ll@{}}
\toprule
\textbf{Distribution} & \textbf{PDF Expression} \\ \midrule
Exponential Decay & 
\( f(x)= d\,\exp\Bigl[-a\,(x-c)\Bigr] \) \\[2ex]
Weibull & 
\( f(x)= d\,c\,(x-b)^{\,c-1}\exp\Bigl[-(x-b)^c\Bigr]\) \\[2ex]
Power Law with Tail (Pareto) & 
\( f(x)= d\,\frac{b}{(x-c)^{\,b+1}}\) \\[2ex]
Cauchy & 
\( f(x)= d\,\frac{2}{\pi}\frac{1}{1+(x-a)^2} \) \\[2ex]
Half Generalized Normal & 
\( f(x)= d\,\frac{b}{\Gamma(1/b)}\,\exp\!\Bigl[-(x-a)^b\Bigr] \) \\[2ex]\\ \bottomrule
\end{tabular}
\vspace{0.5em} 
\begin{minipage}{0.9\linewidth}
\small
\textbf{Notes:}
The parameter $d$ was used as an overall scaling factor.
\end{minipage}
\end{table}

\section{Milky Way 2D-plane map}

\begin{figure}
	\includegraphics{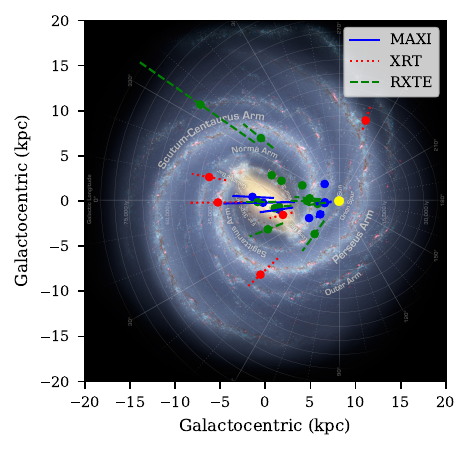}
    \caption{This figure shows the bias-corrected locations of sources in our \citet{Abdulghani2024} original sample. Background image: NASA/JPL-Caltech/R. Hurt (SSC/Caltech).} 
    \label{fig:2D-spatial_distribution}
\end{figure}


\bsp	
\label{lastpage}
\end{document}